\documentclass[11pt]{article}

\usepackage[margin=1in]{geometry} 
\usepackage{amsmath,amssymb,amsthm,booktabs,array}
\usepackage[hidelinks,pdfencoding=auto,psdextra]{hyperref}

\newtheorem{theorem}{Theorem}[section]
\newtheorem{proposition}[theorem]{Proposition}
\newtheorem{lemma}[theorem]{Lemma}
\newtheorem{corollary}[theorem]{Corollary}
\newtheorem{remark}[theorem]{Remark}
\newtheorem{definition}[theorem]{Definition}

\numberwithin{equation}{section}
\numberwithin{table}{section}
\numberwithin{figure}{section}

\newcommand{\F}{\mathbb F}
\newcommand{\C}{\mathbb C}

\newcommand{\Tr}{\operatorname{Tr}}
\newcommand{\rank}{\operatorname{rank}}
\newcommand{\supp}{\operatorname{supp}}
\newcommand{\ket}[1]{\lvert #1\rangle}
\newcommand{\bra}[1]{\langle #1\rvert}
\newcommand{\proj}[1]{\lvert #1\rangle\!\langle #1\rvert}
\newcommand{\AEQ}{\operatorname{EQ}}
\newcommand{\GEQ}{\operatorname{GEQ}}
\newcommand{\MAJ}{\operatorname{MAJ}_3}
\newcommand{\EQ}{\operatorname{EQ}_3}
\newcommand{\PAR}{\operatorname{PAR}_3}
\newcommand{\AND}{\operatorname{AND}_3}

\newcommand{\QPSQMS}{Q^{\mathrm{PSQM},*}}

\title{Concrete Functions in Private Simultaneous Quantum Messages: Quantum Advantages and Structural Limitations}
\author{Harumichi Nishimura\\ Graduate School of Informatics, Nagoya University}
\date{}

\begin{document}

\maketitle

\begin{abstract}
Private simultaneous quantum messages (PSQM) provide a minimal model of noninteractive secure multiparty quantum computation, 
yet their communication complexity for concrete functions remains poorly understood. 
We show that familiar total functions already exhibit both genuine quantum advantages and strong structural limitations.

For three-party equality on \(n\)-bit inputs, which determines whether the players' $n$-bit strings are identical, 
we construct an entanglement-assisted protocol with asymptotic communication \((1.633\ldots)n\) qubits, 
whereas every perfectly correct protocol without prior entanglement requires at least \(3n\) qubits. 
For three-bit majority, one shared GHZ state and one transmitted qubit per player suffice for perfect correctness and perfect privacy, 
reducing the communication from the optimal classical cost of six bits to three qubits.

Three-bit AND behaves differently. 
No perfectly correct and perfectly private protocol exists when the players share an arbitrary three-qubit state, apply arbitrary local unitary encodings, and transmit one qubit each. 
However, increasing the local dimension overcomes this obstruction: one shared three-qutrit GHZ state and one transmitted qutrit per player yield a one-shot protocol with total communication \(3\log_2 3\). Together with the six-bit classical lower bound, this gives a strict quantum communication advantage. More generally, for every \(k\ge3\), we construct a perfectly correct and perfectly private one-shot protocol for \(k\)-bit AND using one qudit of dimension at most \(k\) per player, and hence at most \(k\log_2 k\) qubits in total.
\end{abstract}

\setcounter{tocdepth}{2}
\tableofcontents
\clearpage 

\section{Introduction}\label{sec:introduction}

Private simultaneous messages (PSM)~\cite{FKN94,IK97} form a minimal model of secure multiparty computation. 
Several spatially separated players receive private inputs and send one message each to a referee, without interaction among the players. 
The referee must recover the prescribed function value while learning nothing beyond that value. 
In the quantum version, called private simultaneous quantum messages (PSQM), 
the transmitted messages may be quantum states, and the players may share prior entanglement in addition to classical randomness. 
Kawachi and Nishimura~\cite{KN21} introduced the PSQM model and demonstrated that quantum communication and prior entanglement can reduce communication 
under perfect correctness and perfect privacy. 
Their results include a factor-two separation between PSQM and PSM for a total function extending two-party equality 
and an exponential separation for a partial function. 
More recently, Girish, May, Orshansky, and Waddell~\cite{GirishMayOrshanskyWaddell2026} obtained an exponential separation 
for a partial function in the robust setting which allows correctness and privacy errors. 
These results demonstrate that genuinely quantum resources can provide substantial advantages across both perfect and approximate privacy regimes. 
The behavior of familiar concrete functions, however, remains much less understood.

This issue already appears in the classical PSM model. 
General upper and lower bounds have been developed~\cite{FKN94,BIKK14,AHMS20}, 
but a substantial gap between them remains. 
Various concrete functions have also been studied~\cite{FKN94,IK97}, 
yet matching upper and lower bounds, and hence the optimal communication complexity, are known in only a few cases. 
Shinagawa and Nuida~\cite{ShinagawaNuida2023} advanced this line of research by deriving explicit lower bounds for concrete functions such as AND, majority, and equality. 
In particular, they determined the optimal classical PSM communication complexity for three-input majority and equality, 
while the precise complexity of three-input AND remained unresolved. 
This suggests that even among very small Boolean functions, 
private communication complexity may depend on structural features that are not captured by ordinary circuit complexity 
or by the apparent simplicity of their algebraic representations. 

This paper initiates a parallel investigation of perfectly correct and perfectly private PSQM protocols. 
Our goal is not merely to exhibit another specially designed separation, 
but to understand which types of concrete functions admit an advantage from prior entanglement 
and which structural features can obstruct efficient private quantum protocols. 
Our results reveal a striking contrast among Boolean functions. 
Entanglement provides communication savings for certain equality- and majority-type functions, 
whereas functions with a unique preimage, such as AND, encounter a strong obstruction when each player applies a local unitary encoding to one qubit and transmits that qubit to the referee. 
This contrast suggests that the complexity of a function in PSQM is not determined merely by its algebraic or computational simplicity. 
Rather, the sets of accepting and rejecting inputs and the effect of local input changes on the shared state play a central role. 

Table~\ref{tab:main-results} summarizes the communication bounds established or used in this paper. 
Standalone values denote optimal communication costs, whereas inequalities and intervals denote upper or lower bounds. 
The classical and quantum communication costs are measured in bits and qubits, respectively.

\begin{table}[t] \centering \small \renewcommand{\arraystretch}{1.25} 
\begin{tabular}{ >{\centering\arraybackslash}m{0.20\textwidth} >{\centering\arraybackslash}m{0.34\textwidth} >{\centering\arraybackslash}m{0.34\textwidth} } \toprule Function & Classical PSM & Entanglement-assisted PSQM \\ \midrule 
\(\operatorname{EQ}_{k,n}\) &  \(\geq kn\) bits & \(\displaystyle \le \frac{k}{2h_2(1/k)}\,n + O_k(\sqrt n) \) qubits \\[1mm] 
\(\operatorname{MAJ}_3\) & \(6\) bits \cite{ShinagawaNuida2023} & \( 3\) qubits \\ 
\(\operatorname{EQ}_3\) & \(3\log_2 3\) bits \cite{ShinagawaNuida2023} & \( 3\) qubits \\ 
\(\operatorname{AND}_3\) & \(\displaystyle [6, 3\log_2 5] \) bits \cite{FKN94,ShinagawaNuida2023} & \(\displaystyle \le 3\log_2 3 \) qubits \\[1mm] 
\(\operatorname{AND}_k\) $(k\geq 3)$ & \(\displaystyle \le k\log_2\ell_k \) bits \cite{FKN94} & \(\displaystyle \le k\log_2 p_k \le k\log_2 k \) qubits \\ \bottomrule 
\end{tabular} 
\caption{Classical and entanglement-assisted quantum communication bounds for the functions studied in this paper. Here, \(\ell_k\) is the smallest prime greater than \(k\), while \(p_k\) is the smallest prime satisfying \(k\le2(p_k-1)\). The interval in the $\operatorname{AND}_3$ row gives the currently known classical lower and upper bounds.} \label{tab:main-results} 
\end{table}

Our results are described in more detail below.

\subsection{Our contributions}\label{subsec:con}
 
We develop this viewpoint through four complementary results. 

\paragraph{Entanglement-assisted advantage for multiparty equality.} 

Our first result establishes an entanglement-assisted advantage for a standard predicate, rather than for a function designed specifically to produce a separation. 
For the multiparty equality function on $n$-bit strings, $\operatorname{EQ}_{k,n}(x_1,\ldots,x_k) = \mathbf 1[x_1=\cdots=x_k]$, 
where \(\mathbf 1[P]\) equals \(1\) if the predicate \(P\) is true and \(0\) otherwise, 
we obtain an asymptotic separation between entanglement-assisted PSQM and PSQM without prior entanglement.

Using multicolored sum-free families, we reduce the equality function to the generalized-equality problem studied in \cite{KN21}. 
For three parties, the asymptotically tight tri-colored sum-free construction of Kleinberg, Sawin, and Speyer~\cite{KSS18}, 
completed by Pebody~\cite{Pebody18}, yields a local embedding into \(\mathbb F_2^d\) with $d = \frac{n}{h_2(1/3)} + O(\sqrt n)$,  
where $h_2$ is the binary entropy function. 
More generally, for fixed \(k\), we use the corresponding construction of Lov\'asz and Sauermann~\cite{LS19}. 

For odd \(k\), the protocol in \cite{KN21} incurs a dummy-user penalty. 
Inspired by the odd-party secure-sum construction of Lu, Yao, and Jafar~\cite{LYJ23}, 
we develop a one-shot four-coordinate secure-sum layer that removes this penalty. 
Let \(Q^{\mathrm{PSQM},*}(f)\) denote the minimum communication cost of a perfectly correct and perfectly private entanglement-assisted PSQM protocol for \(f\).
The resulting bound is 
\[ 
Q^{\mathrm{PSQM},*}(\operatorname{EQ}_{k,n}) \leq \frac{k}{2h_2(1/k)}\,n+O_k(\sqrt n). 
\]   
For \(k=3\), this becomes $Q^{\mathrm{PSQM},*}(\operatorname{EQ}_{3,n}) \leq (1.633\ldots)n+O(\sqrt n)$. 
In contrast, every perfectly correct PSQM protocol without prior entanglement, 
even with arbitrary shared classical randomness, requires at least \(kn\) communicated qubits. 
Hence the construction gives a strict asymptotic separation for every fixed \(3\leq k\leq9\). 

\paragraph{Three-bit majority and the local-Pauli landscape.} 

We next study three-party perfectly correct and perfectly private PSQM protocols for three-bit functions in which each party has a one-bit input. 
Starting from the optimal classical six-bit PSM protocol for the three-bit majority \(\operatorname{MAJ}_3\) due to Shinagawa and Nuida~\cite{ShinagawaNuida2023}, 
we obtain a three-qubit protocol for $\MAJ$ using one shared GHZ state, local Pauli encodings, and one transmitted qubit per player. 
We then compress the shared randomness to two bits while preserving the same two output-dependent density operators. 
A local parity twist converts this protocol into one for the three-bit equality \(\operatorname{EQ}_3\). 
These protocols yield quantum communication advantages 
since the optimal classical communication costs of $\MAJ$ and $\EQ$ are $6$ and $3\log_2 3$ bits, respectively \cite{ShinagawaNuida2023}. 

We further construct protocols for a family of exactly \(80\) three-bit Boolean functions. 
Writing the algebraic normal form as 
\[ 
f(x,y,z) = c\oplus ax\oplus by\oplus dz \oplus\alpha xy\oplus\beta xz\oplus\gamma yz \oplus\delta xyz, 
\] 
we construct protocols whenever $\delta=0$ and $(\alpha,\beta,\gamma) \in \{(0,0,0),(1,0,0),(0,1,0),(0,0,1),(1,1,1)\}$.  
These conditions describe exactly 80 Boolean functions. 

\paragraph{A structural barrier for AND and unique preimage.} 

Although the three-bit AND function \(\operatorname{AND}_3\) may appear simpler than \(\operatorname{MAJ}_3\), 
the two functions behave very differently in our PSQM setting. 

Let \(f:\{0,1\}^k\to\{0,1\}\) be a function for which one of the two output values has exactly one preimage. 
We prove that no perfectly correct and perfectly private PSQM protocol exists in the \emph{one-qubit local-unitary model}: 
the players share an arbitrary \(k\)-qubit mixed state and arbitrary classical randomness, apply arbitrary local \(U(2)\) encodings, and transmit one qubit each. 

The proof combines a structural property of one-qubit unitaries with rank bounds. 
After complementing the output if necessary, suppose that the unique input has output \(1\). 
Perfect correctness gives vanishing overlaps between its received state and the states obtained by nonempty sets of input flips. 
For single-coordinate flips, the one-qubit structural lemma (Lemma~\ref{lem:rigidity}) forces the corresponding relative unitaries to be Hermitian involutions up to global phase. 
The full family of input-flip conditions then produces \(2^k-1\) mutually orthogonal pure-state components corresponding to output \(0\) in \(\operatorname{supp}(\varrho_0)\), 
where $\varrho_0$ denotes the output-dependent state received by the referee whenever the function value is \(0\). 
Hence $\operatorname{rank}(\varrho_0)\ge2^k-1$. 
The resulting one-qubit structure, together with perfect privacy, confines this support to a three-dimensional local-action subspace, giving $\operatorname{rank}(\varrho_0)\le3$. 
These bounds are incompatible for \(k\ge3\).

The theorem applies to all Boolean point functions, including the $k$-bit AND function $\operatorname{AND}_k$, and their complements. 
The contrast between \(\operatorname{MAJ}_3\) and \(\operatorname{AND}_3\) therefore illustrates two different roles of entanglement in PSQM: a single GHZ state yields an efficient protocol for majority, whereas no shared three-qubit state suffices for AND in the one-qubit local-unitary model.

\paragraph{One-qudit protocols for multiparty AND.} 

The one-qubit obstruction is not an impossibility for PSQM itself: it can be overcome by increasing the local message dimension. 
We construct a one-shot PSQM protocol for \(\operatorname{AND}_k\) using a single generalized GHZ state. 
A zero-sum-free construction ensures that the all-one input is distinguished from every other input, 
and a finite-field randomization hides which rejecting input was given. 
For every prime \(p\) satisfying $k\le2(p-1)$, the protocol uses one \(p\)-dimensional qudit per player and has communication cost \(k\log_2 p\). 
Since such a prime can always be chosen with \(p\le k\) by Bertrand's postulate, we obtain  
\[
 Q^{\mathrm{PSQM},*}(\operatorname{AND}_k) \le k\log_2 k. 
\] 
For \(k=3\), the construction gives a one-qutrit protocol with communication \(3\log_2 3\). 
Together with the classical lower bound of \(6\) bits in~\cite{ShinagawaNuida2023}, 
this yields a strict quantum communication advantage for three-bit AND.

\subsection{Related work}

\paragraph{Classical private simultaneous messages for concrete functions.} 

Shinagawa and Nuida~\cite{ShinagawaNuida2023} derived explicit upper and lower bounds for several concrete functions. 
Their protocols and lower bounds, including the optimal bounds for \(\operatorname{MAJ}_3\) and $\EQ$, provide the principal classical comparison for our three-bit protocols. 
For two-input AND, the protocol of Feige, Kilian, and Naor~\cite{FKN94} and the matching lower bound of Data, Prabhakaran, and Prabhakaran~\cite{DPP16} establish the optimal classical communication cost of $2\log_2 3$ bits.

\paragraph{Quantum versions of private simultaneous messages.} 

Several quantum variants of the private simultaneous-message model have been studied. 
In PSQM, the model considered in this paper, the players receive classical inputs and may transmit quantum messages. 
In quantum private simultaneous messages (QPSM), each player may instead receive a quantum input register and transmit a quantum message to the referee.

In~\cite{KN21}, the authors introduced PSQM and established separations between entanglement-assisted and shared-randomness-only protocols. 
Christensen and Popovski~\cite{ChristensenPopovski2023} constructed private-product protocols over prime fields using shared entanglement. 
Over the binary field, their protocol computes two-input AND using a shared Bell pair and one transmitted qubit from each player. 
Together with the classical bound above, this yields an entanglement-assisted communication advantage for \(\operatorname{AND}_2\). 
Our impossibility theorem contrasts this positive two-party result with the one-qubit local-unitary obstruction for \(\operatorname{AND}_3\). 

Allerstorfer, Buhrman, May, Speelman, and Verduyn Lunel~\cite{ABMSV24} related PSQM to nonlocal quantum computation 
by deriving PSQM protocols from coherent function evaluation, a special case of position verification. 
Girish, May, Parham, and Yuen~\cite{GMPY25} showed that 
simultaneous quantum-message protocols with a low-\(T\)-depth referee can be transformed into entanglement-assisted PSM protocols with {\em classical} messages and approximate privacy. 
In subsequent work~\cite{GMPY26}, they established general upper and lower bounds for PSM and PSQM, 
including correlation-cost lower bounds based on Ne\v{c}iporuk's measure and communication-matrix rank, and upper bounds based on Clifford $+$ $T$-gates decomposition and Fourier \(1\)-norm. 
These works provide general constructions, resource bounds, and connections with other computation models and complexity measures. 
Our results are complementary in focusing on perfect correctness, perfect privacy, and the communication complexity of concrete functions. 

Kawachi and Nishikubo~\cite{KN25} introduced private quantum signal processing 
and obtained PSM protocols with a one-qubit quantum evaluator for symmetric Boolean functions and functions admitting efficient one-qubit programs. 
These protocols send classical gate descriptions and therefore operate in a different resource regime from ours. 

\paragraph{Quantum decomposable randomized encodings and QPSM.} 

In the classical setting, decomposable randomized encodings correspond to PSM protocols under an appropriate partition of the input. 
Brakerski and Yuen~\cite{BY22} developed a quantum analogue by introducing quantum randomized encodings 
and constructing a quantum decomposable randomized encoding (QDRE) for general quantum circuits. 
Gur, Huang, and Tang~\cite{GurHuangTang2026} subsequently established a corresponding equivalence between QDRE and QPSM for the qubit-by-qubit case: 
QDREs for \(k\)-qubit inputs decomposed into individual input qubits are equivalent to \(k\)-party QPSM protocols with one-qubit local inputs, 
with preservation of quantum encoding size and quantum communication. 

This correspondence does not directly cover our model. Their QPSM setting has quantum inputs and charges only quantum communication, 
whereas in our PSQM setting both classical and quantum message components contribute to the total communication cost.

\subsection{Organization} 
Section~2 defines the PSQM model and gives the required linear-algebraic preliminaries. 
Section~3 proves the entanglement-assisted upper bounds for $\AEQ_{k,n}$. 
Section~4 gives the PSQM protocols for $\MAJ$ and $\EQ$, the analytic construction of 80 Boolean functions, and the finite linear-programming experiment. 
Section 5 proves the unique-preimage impossibility theorem in the one-qubit local-unitary model. 
Section 6 develops the PSQM protocol for multiparty AND while Appendix~\ref{app:qutrit-randomness-compression} gives a three-party-specific shared-randomness compression. 
Section 7 discusses the resulting picture for concrete functions and open problems.

\section{Preliminaries}

\subsection{Classical private simultaneous messages} \label{subsec:psm-model} 

We first recall a classical private simultaneous-message (PSM) protocol. 
Let \(k\geq 2\), let $X = X_1\times\cdots\times X_k$, and let $f:X\longrightarrow Y$ be a function with finite input and output alphabets. 
We call the inputs $x$ with $f(x)=1$ and $f(x)=0$ the accepting inputs and the rejecting inputs, respectively. 

In a PSM protocol, player \(P_i\) receives its local input \(x_i\in X_i\), 
and the players share an input-independent classical random variable \(r\) 
whose distribution $p$ has finite support \(\mathcal R\),\footnote{We restrict the shared randomness to distributions with finite support. 
This entails no loss of generality in the finite-input, finite-dimensional setting considered here.} 
and \(r\in\mathcal R\) is hidden from the referee. 
Player \(P_i\) sends $m_i(x_i;r)\in\mathcal M_i$ to the referee, 
and the referee applies a decoding function $D:\mathcal M_1\times\cdots\times\mathcal M_k\longrightarrow Y$ to obtain the output. 
The players send their messages simultaneously and do not communicate with one another during the protocol. 

Perfect correctness requires 
\[
 D\bigl( m_1(x_1;r),\ldots,m_k(x_k;r) \bigr) = f(x_1,\ldots,x_k) 
\] 
for every input \(x=(x_1,\ldots,x_k)\) and every \(r\in\mathcal R\). 
For each input \(x\), let \(D_x\) denote the distribution of the joint message tuple 
\[
 \bigl( m_1(x_1;r),\ldots,m_k(x_k;r) \bigr), 
\] 
where the probability is over the shared randomness. 

Perfect privacy requires 
\[
 f(x)=f(x') \quad\Longrightarrow\quad D_x=D_{x'}. 
\] 
Thus the referee's complete view reveals no information about the input beyond the value of \(f\). 

The communication cost is $\sum_{i=1}^k\log_2|\mathcal M_i| = \log_2\left(\prod_{i=1}^k|\mathcal M_i|\right)$.  
A protocol with message-alphabet sizes $|\mathcal M_1|=d_1,\ldots,|\mathcal M_k|=d_k$ is called a $(d_1,\ldots,d_k)$-PSM protocol.  
For example, a \((4,4,4)\) PSM protocol has three players, each of whom sends one of four possible messages. 
Its total communication cost is $\log_2 4+\log_2 4+\log_2 4 = 6$ bits. 

\subsection{Private simultaneous quantum messages} \label{subsec:psqm-model} 

We assume familiarity with the basics of quantum information and quantum computation and standard notations such as Pauli matrices~\cite{NC00}. 
$\mathsf L(\mathcal A)$ denotes the set of all linear operators on a linear space $\mathcal A$, 
and $\mathsf D(\mathcal B)$ denotes the set of mixed states over a quantum register $\mathcal B$. 

The private simultaneous quantum message (PSQM) model replaces the classical message alphabets by quantum message registers 
and additionally allows the players to share prior entanglement. 

We now give the precise PSQM model used throughout this paper. 
This is an equivalent reformulation of the original PSQM model in~\cite{KN21}, 
stated directly in terms of the joint message state received by the referee. 
Whereas the original formulation expresses correctness through the referee's output probability and privacy through the existence of a simulator, 
we equivalently formulate these conditions below as properties of the averaged joint message state.

Let \(k\geq 2\), let $X = X_1\times\cdots\times X_k$, and let $f:X\longrightarrow Y$ be a function with finite input and output alphabets. 
An input to the protocol is a tuple $x=(x_1,\ldots,x_k)\in X$, where player \(P_i\) receives only its local input \(x_i\in X_i\). 
A \(k\)-party PSQM protocol for \(f\) consists of the following objects. 

\begin{enumerate} 
\item An input-independent shared random variable \(r\) whose distribution $p$ has finite support \(\mathcal R\). 
\item An input-independent multipartite state $\sigma_{\mathcal A_1\cdots \mathcal A_k} \in \mathsf D(\mathcal A_1\otimes\cdots\otimes \mathcal A_k)$, shared among the players and independent of \(r\). Player \(P_i\) holds register \(\mathcal A_i\). The referee holds no part of these setup resources before the communication.
\item For every player \(P_i\), local input \(x_i\in X_i\), and \(r\in\mathcal R\), 
a quantum channel $\mathcal E_i^{x_i,r}: \mathsf L(\mathcal A_i)\longrightarrow\mathsf L(\mathcal M_i)$.  
Player \(P_i\) applies \(\mathcal E_i^{x_i,r}\) to register \(\mathcal A_i\) and sends the resulting message register \(\mathcal M_i\) to the referee. 
\item A decoding measurement, represented by a POVM $\{D_y : y\in Y\}$, on the joint message register $\mathcal M:=\mathcal M_1\otimes\cdots\otimes \mathcal M_k$. 
\end{enumerate}

The players send their messages simultaneously and do not communicate with one another during the protocol. 
For a global input \(x=(x_1,\ldots,x_k)\) and a fixed value \(r\) of shared randomness, 
define 
\[ 
\rho_{x,r} := \left( \mathcal E_1^{x_1,r}\otimes\cdots\otimes \mathcal E_k^{x_k,r} \right)(\sigma). 
\] 
Since the shared randomness is hidden from the referee, the referee actually receives 
\[ 
\rho_x := \sum_{r\in\mathcal R}p(r)\rho_{x,r}. 
\]
Thus, throughout this paper, \(x\) denotes the global input tuple, 
whereas \(\rho_x\) denotes the joint state of all message registers received by the referee on input \(x\), 
after averaging over all input-independent shared randomness. 

The protocol is \emph{perfectly correct} if, for every \(x\in X\), 
\[
 \operatorname{Tr}\!\left(D_{f(x)}\rho_x\right)=1. 
\] 
The protocol is \emph{perfectly private} if there exist density operators $\{\varrho_y\in\mathsf D(\mathcal M):y\in Y\}$ such that $\rho_x=\varrho_{f(x)}$ for every \(x\in X\). 
Equivalently, 
\[
 f(x)=f(x') \quad\Longrightarrow\quad \rho_x=\rho_{x'}. 
\] 
This is the simulation-based privacy condition for a classical-output function: 
a simulator given only \(y=f(x)\) can reproduce the referee's global view by preparing \(\varrho_y\). 

If the protocol is both perfectly correct and perfectly private, then the output-dependent states have mutually orthogonal supports: 
\[ \operatorname{supp}(\varrho_y) \perp \operatorname{supp}(\varrho_{y'}) \qquad (y\neq y'). \]

The communication cost of the protocol is 
\begin{equation} 
\sum_{i=1}^k\log_2\dim \mathcal M_i = \log_2\dim \mathcal M. \label{eq:cost} 
\end{equation}
No additional free classical transcript is allowed. 
If a message contains a classical string, that string is represented by an orthogonal register and its dimension is included in~\eqref{eq:cost}.

A protocol is {\em shared-randomness-only} if the shared state is a product state, 
\[
 \sigma_{\mathcal A_1\cdots \mathcal A_k} = \sigma_{\mathcal A_1}\otimes\cdots\otimes\sigma_{\mathcal A_k}. 
\] 
Since the input-independent local states \(\sigma_{\mathcal A_1},\ldots,\sigma_{\mathcal A_k}\) can be absorbed into the local encoding channels, 
this is equivalent to the usual model in which the players share only the classical random variable.

An {\em entanglement-assisted} protocol may use an arbitrary multipartite mixed state $\sigma_{\mathcal A_1\cdots \mathcal A_k}$, independent of the shared random variable. 
The shared state and the shared randomness are input-independent and are not charged as communication. 
We nevertheless specify these resources explicitly when stating concrete protocols.

We write $Q^{\mathrm{PSQM}}(f)$ for the minimum communication cost of a perfectly correct and perfectly private shared-randomness-only protocol for \(f\), 
and $Q^{\mathrm{PSQM},*}(f)$ for the corresponding minimum cost of an entanglement-assisted protocol. 

\begin{remark}[Relation to the original PSQM model] \label{rem:original-psqm} 
The definition above is equivalent to the PSQM model in~\cite{KN21}. 
We describe a fixed shared state and independent shared randomness simultaneously, 
since the latter can equivalently be represented by correlated orthogonal registers in the shared state. 
Our main difference is presentational: we formulate correctness and privacy directly in terms of the averaged joint message state \(\rho_x\).
\end{remark}

\begin{remark}[Relation to decomposable randomized encodings] \label{rem:dre} 
For classical inputs and a classical output, a PSQM protocol in our model may be viewed as an information-theoretic quantum decomposable randomized encoding 
with respect to the fixed input partition $X=X_1\times\cdots\times X_k$,  
whose setup consists of a fixed multipartite state together with independent shared classical randomness. 
The collection of message registers generated by the local encoders forms the decomposable encoding, 
the referee's measurement is the decoder, and the output-dependent states in the privacy condition give the simulator.
Conversely, a quantum decomposable randomized encoding with this input partition and 
with a setup consisting of independent shared classical randomness and a fixed shared quantum state gives a PSQM protocol in our model. 
Under this correspondence, the total encoding size equals the PSQM communication cost.
\end{remark}

\subsection{Two linear-algebra lemmas}

\begin{lemma}[support of a positive mixture]\label{lem:support}
If $A=\sum_jc_jA_j$ with $c_j>0$ and $A_j\geq0$, then $\supp(A_j)\subseteq\supp(A)$ for every $j$.  Hence, if $A$ and $B$ have orthogonal supports, every positive component of $A$ has support orthogonal to every positive component of $B$.
\end{lemma}

\begin{proof}
If $\ket z\in\ker A$, then $0=\sum_jc_j\bra zA_j\ket z$.  
Since every term is nonnegative, $A_j\ket z=0$ for every $j$.  
Thus $\ker A\subseteq\ker A_j$, equivalently $\supp(A_j)\subseteq\supp(A)$.
\end{proof}

\begin{lemma}[zero-expectation qubit-unitary structure]\label{lem:rigidity}
Let $\tau$ be a one-qubit density operator and $V\in U(2)$.  
If $\Tr(\tau V)=0$, then up to a global phase, $V$ is a Hermitian involution: $V^{\dagger}=V$ and $V^2=I$.
\end{lemma}

\begin{proof} 
Let $\boldsymbol{\sigma} := (X,Y,Z)$ denote the vector of Pauli matrices. 
For \(\boldsymbol v=(v_x,v_y,v_z)\in\mathbb R^3\), 
write 
\[ 
\boldsymbol v\cdot\boldsymbol{\sigma} := v_x X+v_y Y+v_z Z. 
\] 
Using the Bloch representation of a qubit state and the standard parametrization of a \(2\times2\) unitary, write 
\[ 
\tau = \frac12 \left( I+\boldsymbol t\cdot\boldsymbol{\sigma} \right), \qquad V = e^{i\alpha} \left( \cos\theta\,I +i\sin\theta\, \boldsymbol n\cdot\boldsymbol{\sigma} \right), 
\] 
where 
\[ 
\boldsymbol t\in\mathbb R^3, \qquad \|\boldsymbol t\|\leq1, \qquad \boldsymbol n\in\mathbb R^3, \qquad \|\boldsymbol n\|=1. 
\]
Let $V'=e^{-i\alpha}V$. 
The direct calculation using standard relations on Pauli matrices (such as Equation (2.78) in \cite{NC00}) gives 
\[ 
\operatorname{Tr}(\tau V') = \cos\theta +i\sin\theta\, \boldsymbol t\cdot\boldsymbol n. 
\] 
The hypothesis \(\operatorname{Tr}(\tau V)=0\) implies \(\operatorname{Tr}(\tau V')=0\) and, by taking the real part, that $\cos\theta=0$. 
Hence \(\sin\theta=\pm1\). 
Let $V''=-i\sin\theta V'=\boldsymbol n\cdot\boldsymbol{\sigma}$.   
Then $V''$ is Hermitian and satisfies $(\boldsymbol n\cdot\boldsymbol{\sigma})^2=I$. 
$V''$ is equal to $V$ up to a global phase.
\end{proof}

\section{Entanglement-Assisted Advantage for Equality on Strings}\label{sec:aeq}

We begin in Section~\ref{subsec:comb-embed} with the combinatorial reduction from $\AEQ_{k,n}$ to generalized equality, 
and recall the previously known protocol, to derive some non-trivial upper bound for $\AEQ_{k,n}$. 
Sections~\ref{subsec:bell-primitive} and~\ref{subsec:odd-sum} develop the Bell-based secure-sum primitives needed to remove the odd-party dummy-user penalty. 
Section~\ref{subsec:private-GEQ} applies these primitives blockwise, and uses random nonzero scaling to obtain a PSQM protocol for generalized equality. 
Section~\ref{subsec:all-EQ-main} combines this protocol with the combinatorial reduction to derive the upper bound for $\AEQ_{k,n}$ claimed in Section~\ref{sec:introduction}.

\subsection{Combinatorial embeddings by multicolored sum-free families}\label{subsec:comb-embed}

For a positive integer $L$, write $[L]=\{1,\ldots,L\}$. 
Let $h_2$ denote the binary entropy function: $h_2(p)=-p\log_2p-(1-p)\log_2(1-p)$. 

\begin{definition}
A $k$-colored sum-free family of size $L$ in an abelian group $G$ is a collection
\[
 \{(a_{1,v},\ldots,a_{k,v}):v\in[L]\}\subseteq G^k
\]
such that
\[
 a_{1,v_1}+\cdots+a_{k,v_k}=0
 \quad\Longleftrightarrow\quad
 v_1=\cdots=v_k.
\]
\end{definition}
For an abelian group $G$, let $\operatorname{sfs}_k(G)$ denote the maximum size of a $k$-colored sum-free family in $G$, 
where $G$ is regarded as an additive group.

For \(k=3\), the corresponding asymptotically tight construction was obtained by Kleinberg, Sawin, and Speyer~\cite{KSS18}, 
together with Pebody's proof of the required conjecture~\cite{Pebody18}. 
Lov\'asz and Sauermann~\cite{LS19} proved a general lower bound for \(k\)-colored sum-free families in \(\mathbb Z_m^d\). 
Applying their Theorem~1.3 to the binary group $\mathbb Z_2^d=\mathbb F_2^d$ gives 
\[
 \operatorname{sfs}_k(\mathbb F_2^d) \ge \Gamma_{2,k}^{\,d-O_k(\sqrt d)}, 
\] 
where 
\[
 \Gamma_{2,k} = \min_{0<\gamma<1} (1+\gamma)\gamma^{-1/k}. 
\] 
Indeed, logarithmic differentiation gives
\[
 \frac{d}{d\gamma} \log\bigl((1+\gamma)\gamma^{-1/k}\bigr) = \frac{(k-1)\gamma-1} {k\gamma(1+\gamma)}. 
\] 
Hence the objective decreases before \(\gamma=1/(k-1)\) and increases thereafter, 
and its unique minimum is attained at \(\gamma=1/(k-1)\). 
It follows that 
\[
 \Gamma_{2,k} = \frac{k}{(k-1)^{(k-1)/k}} = 2^{h_2(1/k)}. 
\] 
Therefore, 
\begin{equation}\label{eq:sum-free}
 \operatorname{sfs}_k(\mathbb F_2^d) \geq 2^{h_2(1/k)d-O_k(\sqrt d)}.
\end{equation}

\begin{lemma}[fixed-\(k\) local embedding]
\label{lem:AEQkembed-revised}
Fix \(k\ge3\). For all sufficiently large \(n\), there are an integer 
\[
 d = \frac{n}{h_2(1/k)} + O_k(\sqrt n) 
\] 
and local maps 
\[
 A_j:\{0,1\}^n\longrightarrow\mathbb F_2^d, \qquad j\in[k], 
\] such that 
\[
 \sum_{j=1}^k A_j(x_j)=0 \quad\Longleftrightarrow\quad x_1=\cdots=x_k. 
\] 
\end{lemma}

\begin{proof}
By~\eqref{eq:sum-free}, 
we may choose $d = \frac{n}{h_2(1/k)} + O_k(\sqrt n)$ such that $\operatorname{sfs}_k(\mathbb F_2^d)\ge2^n$. 
Choose a \(k\)-colored sum-free family of size \(2^n\), and index its members by $x\in\{0,1\}^n$:
\[
 \left\{ (a_{1,x},\ldots,a_{k,x}):x\in \{0,1\}^n \right\} \subseteq (\mathbb F_2^d)^k.  
\] 
For each \(j\in[k]\), define $A_j(x):=a_{j,x}$. 
Then 
\[
 \sum_{j=1}^k A_j(x_j)=0 \quad\Longleftrightarrow\quad x_1=\cdots=x_k, 
\] 
by the defining property of the sum-free family.
\end{proof}

The sum-free theorem uses only the additive group. 
In particular, $\mathbb Z_2^d=(\mathbb F_2^d,+)$. 
Field multiplication is introduced only after the embedding, 
through a chosen additive isomorphism with \(\mathbb F_{2^d}\). 
No multiplicative closure of the sum-free family is assumed or required.

For $u_1,\ldots,u_k\in\F_2^d$, define the generalized equality function\footnote{This name was used in~\cite{KN21} 
since it matches the equality function for the $k=2$ case.}  
\[
 \GEQ_{k,d}(u_1,\ldots,u_k)
 =\mathbf1[u_1+\cdots+u_k=0].
\]
In~\cite{KN21}, the authors gave perfectly correct and perfectly private entanglement-assisted protocols satisfying, 
for even $d$,
\begin{equation}
 \QPSQMS(\GEQ_{k,d})
 \leq
 \begin{cases}
 kd/2,&k\text{ even},\\
 (k+1)d/2,&k\text{ odd}.
 \end{cases}
 \label{eq:KNoriginal-revised}
\end{equation}
Note that for odd \(k\), the last party simulates an additional zero-input dummy party. 

Combining Lemma~\ref{lem:AEQkembed-revised} with \eqref{eq:KNoriginal-revised}, 
and padding the embedded vectors by at most one zero coordinate, 
gives the parity-dependent bounds 
\[
 Q^{\mathrm{PSQM},*}(\AEQ_{k,n}) \le \begin{cases} \displaystyle \frac{k}{2h_2(1/k)}\,n + O_k(\sqrt n), & k \text{ even},\\[3mm] \displaystyle \frac{k+1}{2h_2(1/k)}\,n + O_k(\sqrt n), & k \text{ odd}. \end{cases} 
\] 
In particular, for \(k=3\), \[ Q^{\mathrm{PSQM},*}(\AEQ_{3,n}) \le \frac{2}{h_2(1/3)}\,n + O(\sqrt n) = (2.177\ldots)n + O(\sqrt n). \] 
We retain these bounds for comparison with the improved odd-party layer below.

\subsection{A binary physical two-user Bell primitive}\label{subsec:bell-primitive}

Let
\[
 \ket{\Phi_{0,0}}=\frac{\ket{00}+\ket{11}}{\sqrt2},
 \qquad
 \ket{\Phi_{a,b}}=(X^aZ^b\otimes I)\ket{\Phi_{0,0}},
 \qquad a,b\in\mathbb F_2.
\]
These four Bell states form an orthonormal basis. 
Since
\[
 (I\otimes M)\ket{\Phi_{0,0}}=(M^T\otimes I)\ket{\Phi_{0,0}},
\]
suitable local Pauli operations map two users' input pairs
\((a_1,a_2)\) and \((b_1,b_2)\) to
\[
 \ket{\Phi_{a_1+b_1,a_2+b_2}}
\]
up to global phase. 
Concretely, the two users initially share $|\Phi_{0,0}\rangle$ with each user holding one qubit, 
and apply \(X^{a_1}Z^{a_2}\) and \(X^{b_1}Z^{b_2}\), respectively.  
Up to a global phase, the referee receives $\lvert\Phi_{a_1+b_1,a_2+b_2}\rangle$ 
and obtains $(a_1+b_1,a_2+b_2)$ by the Bell-basis measurement. 
Thus two transmitted qubits carry two binary sums, 
and the joint state received by the referee depends only on those sums. 

In what follows, we call this protocol the two-user Bell primitive.

\subsection{Odd-party four-coordinate secure sum}\label{subsec:odd-sum}

\begin{lemma}[four-coordinate secure sum for odd \(k\)]
\label{lem:oddsecure}
Let \(k\geq3\) be odd. Player \(i\), for $i=1,\ldots,k$, holds four bits
\[
 w_i^{(1)},w_i^{(2)},w_i^{(3)},w_i^{(4)}\in\mathbb F_2.
\]
There is a perfectly correct and perfectly private PSQM protocol for the four sums
\[
 S^{(\ell)}=\sum_{i=1}^k w_i^{(\ell)}\in\mathbb F_2,
 \qquad \ell=1,2,3,4.
\]
The protocol uses exactly \(2k\) transmitted qubits and no dummy party.
\end{lemma}

\begin{proof}
We use players \(1,2,3\) to run two independent triangle instances. The first instance handles the coordinate pair \((1,2)\), and the second handles the coordinate pair \((3,4)\). Each triangle instance assigns one binary value to each of the three edges \[ \{1,2\},\qquad \{1,3\},\qquad \{2,3\}. \] We will pack the two values assigned to the same edge into one application of the two-user Bell primitive from Section~\ref{subsec:bell-primitive}.

We first describe one triangle instance. 
Suppose that players \(1,2,3\) hold input pairs \[ (u_i,v_i)\in\mathbb F_2^2, \qquad i=1,2,3. \] 
Choose a uniform \(r\in\mathbb F_2\), and assign the edge values 
\[
 \begin{aligned} Y_{12}&=v_1+(u_2+r),\\ Y_{13}&=(u_1+v_1)+(u_3+r),\\ Y_{23}&=(u_2+v_2)+(v_3+r) \end{aligned} 
\] to the edges $\{1,2\},\{1,3\},\{2,3\}$, respectively. 
Each edge value is the sum of one bit locally computable by each endpoint. 
For example, on edge \(\{1,2\}\), player \(1\) contributes \(v_1\), while player \(2\) contributes \(u_2+r\). 
Similarly, on edge \(\{1,3\}\), players \(1\) and \(3\) contribute \(u_1+v_1\) and \(u_3+r\), respectively; on edge \(\{2,3\}\), players \(2\) and \(3\) contribute \(v_2+u_2\) and \(v_3+r\), respectively. Since \(r\) is shared randomness, all these contributions are locally computable by the indicated endpoints.

The referee can reconstruct 
\[
 Y_{12}+Y_{13} = u_1+u_2+u_3, \qquad Y_{12}+Y_{23} = v_1+v_2+v_3. 
\] 
Let \[ S_u:=\sum_{i=1}^3u_i, \qquad S_v:=\sum_{i=1}^3v_i. \] Setting \(t:=Y_{12}\), we have \[ (Y_{12},Y_{13},Y_{23}) = (t,S_u+t,S_v+t). \] 
Since \(r\) is uniform, so is \(t\). Thus, conditioned on \((S_u,S_v)\), the joint distribution of the three edge values is uniform over 
\[ \left\{ (t,S_u+t,S_v+t): t\in\mathbb F_2 \right\}, \] and hence depends only on \((S_u,S_v)\).

Apply this construction first with \[ (u_i,v_i) = \bigl(w_i^{(1)},w_i^{(2)}\bigr) \] and independently with \[ (\widetilde u_i,\widetilde v_i) = \bigl(w_i^{(3)},w_i^{(4)}\bigr). \] 
Denote the resulting edge values by 
\[ Y_{ab}^{[1]} \quad\text{and}\quad Y_{ab}^{[2]}, \qquad \{a,b\} \in \bigl\{ \{1,2\},\{1,3\},\{2,3\} \bigr\}. \] 
For each edge \(\{a,b\}\), its endpoints now have two pairs of local contributions, one pair for each triangle instance. They use these as their two-bit inputs to one application of the Bell primitive in Section~\ref{subsec:bell-primitive}. The referee consequently recovers $\bigl(Y_{ab}^{[1]},Y_{ab}^{[2]}\bigr)$ by the Bell-basis measurement. 
Thus each edge uses one Bell pair, and the two triangle instances together use three Bell pairs and transmit six qubits.
 
Now write \(k=2m+1\) and partition the remaining players into the unordered pairs 
\[
 \{4,5\},\{6,7\},\ldots,\{k-1,k\}, 
\] 
with the collection understood to be empty when \(k=3\). 
Using the common classical randomness shared by all \(k\) players, choose independent uniform masks 
\[
 z_j^{(\ell)}\in\mathbb F_2, \qquad j=1,\ldots,m-1, \qquad \ell=1,2,3,4. 
\] 
For pair \(j\), denote its endpoints by \(a_j\) and \(b_j\), and let the pair supply 
\[
 Q_j^{(\ell)} = w_{a_j}^{(\ell)} + w_{b_j}^{(\ell)} + z_j^{(\ell)}. 
\] 
Each remaining pair shares two Bell pairs, one for coordinates \((1,2)\) and one for coordinates \((3,4)\), 
and therefore transmits four qubits. 
The first three users run the two triangle instances with shifted partial sums 
\[
 T^{(\ell)} = w_1^{(\ell)} + w_2^{(\ell)} + w_3^{(\ell)} + \sum_{j=1}^{m-1}z_j^{(\ell)}. 
\] 
For each coordinate \(\ell\), the shift $\sum_{j=1}^{m-1}z_j^{(\ell)}$ is implemented locally 
by adding this shared offset to one selected input of the corresponding three-user triangle. 
The referee then reconstructs 
\[ 
\begin{aligned} 
T^{(\ell)} + \sum_{j=1}^{m-1}Q_j^{(\ell)} 
&= w_1^{(\ell)} + w_2^{(\ell)} + w_3^{(\ell)} + \sum_j z_j^{(\ell)}+ \sum_j \left( w_{a_j}^{(\ell)} + w_{b_j}^{(\ell)} + z_j^{(\ell)} \right)\\ 
&= S^{(\ell)}. 
\end{aligned} 
\] 
We next verify privacy. 
For fixed inputs, the random variables $Q_j^{(\ell)}$ are independent uniform bits, 
because the masks \(z_j^{(\ell)}\) are independent and uniform. Conditioned on the four global sums, 
the shifted triangle targets are determined by 
\[
 T^{(\ell)} = S^{(\ell)} + \sum_{j=1}^{m-1}Q_j^{(\ell)}. 
\] 
For the triangle instance associated with coordinates \((\ell_1,\ell_2)\), 
conditioned on $\left( T^{(\ell_1)},T^{(\ell_2)} \right)$, 
its three edge values are uniform over 
\[
 \left\{ \left( t, T^{(\ell_1)} + t, T^{(\ell_2)} + t \right): t\in\mathbb F_2 \right\}. 
\] 
It follows that the joint distribution of all edge values depends only on the four global sums 
\[
 \left( S^{(1)},S^{(2)},S^{(3)},S^{(4)} \right). 
\] 
Each Bell block is, up to global phase, 
a pure orthogonal Bell state completely determined by its two edge values. 
Therefore the averaged joint state received by the referee also depends only on the four global sums. 
Finally, the first three players transmit six qubits, and each of the remaining $\frac{k-3}{2}=m-1$ pairs transmits four qubits. 
Hence the total communication is $6 + 4\frac{k-3}{2} = 2k$ qubits.
\end{proof}

The triangle, pairing, and mask-cancellation pattern are based on the odd-\(k\) secure-sum idea of Lu, Yao, and Jafar~\cite{LYJ23}. 
Lemma~\ref{lem:oddsecure} expands their normalized additive-channel accounting into physical Bell-state primitives and makes explicit the finite four-coordinate block used in our one-shot generalized-equality protocol.\footnote{``one-shot'' means that one joint input tuple produces one function value. 
The block coding is internal to that tuple and is not amortization over independent instances.}

\subsection{From blockwise secure sums to private generalized equality}\label{subsec:private-GEQ}

The preceding subsection gives a four-coordinate secure-sum block for odd \(k\), 
while the protocol of~\cite{KN21} gives the required two-coordinate secure-sum block for even \(k\). 
After padding and partitioning the embedded vectors into blocks of the appropriate size, 
these primitives produce a received state that depends only on the full aggregate $s=\sum_{i=1}^k y_i$. 
Correctness for generalized equality requires distinguishing \(s=0\) from \(s\ne0\), 
but perfect privacy additionally requires all nonzero aggregates to induce the same received state. 
We enforce this by multiplying all embedded inputs by the same uniformly random nonzero element of an extension field.

Let \(d'\) be the padded coordinate length: a multiple of two for even \(k\), and a multiple of four for odd \(k\). 
Embed each original input \(y_i\in\mathbb F_2^d\) as $\widetilde y_i = (y_i,0^{d'-d}) \in\mathbb F_2^{d'}$.  
Its aggregate $\widetilde s = \sum_{i=1}^k \widetilde y_i$ vanishes if and only if the original aggregate $s = \sum_{i=1}^k y_i$ vanishes. 
For notational simplicity, 
we subsequently write \(y_i\) and \(s\) for the padded vectors \(\widetilde y_i\) and \(\widetilde s\), respectively. 
For every aggregate \(s\in\mathbb F_2^{d'}\), 
let \(\Omega_s\) denote the global density operator received from the secure-sum layer when the aggregate vector is \(s\). 
This notation is well defined: if two tuples $(y_1,\ldots,y_k)$ and $(y'_1,\ldots,y'_k)$ satisfy \[ \sum_{i=1}^k y_i = \sum_{i=1}^k y'_i = s, \] 
then the aggregate-privacy property of the secure-sum layer 
(namely, the $Sum_2$ protocol of \cite{KN21} for even $k$ and Lemma~\ref{lem:oddsecure} for odd $k$) implies that 
the two tuples produce exactly the same received density operator \(\Omega_s\). 
Fix an \(\mathbb F_2\)-linear isomorphism $\iota: \mathbb F_2^{d'} \longrightarrow (\mathbb F_{2^{d'}},+)$. 

The players choose a shared uniform \(\lambda\in\mathbb F_{2^{d'}}^\times\) and locally replace each input
vector \(y_i\) by $\iota^{-1}(\lambda\,\iota(y_i))$. 
This is an \(\mathbb F_2\)-linear operation, and therefore the new aggregate is $s_\lambda =\iota^{-1}(\lambda\,\iota(s))$. 
After averaging over \(\lambda\), the received global density operator is
\begin{equation}
 \overline\Omega_s
 =\frac{1}{2^{d'}-1}
 \sum_{\lambda\in\mathbb F_{2^{d'}}^\times}
 \Omega_{\iota^{-1}(\lambda\iota(s))}.
 \label{eq:density-twirl-revised}
\end{equation}
Consequently,
\begin{equation}
 \overline\Omega_s
 =
 \begin{cases}
 \Omega_0,&s=0,\\[5pt]
 \displaystyle
 \frac{1}{2^{d'}-1}
 \sum_{t\in\mathbb F_2^{d'}\setminus\{0\}}\Omega_t,
 &s\neq0.
 \end{cases}
 \label{eq:zero-nonzero-states-revised}
\end{equation}
Indeed, if \(s\neq0\), multiplication by \(\iota(s)\) permutes
\(\mathbb F_{2^{d'}}^\times\).  
Equation \eqref{eq:zero-nonzero-states-revised} proves perfect Boolean privacy at the level of the received global density operator: 
all zero aggregates give one fixed state, and all nonzero aggregates give a second fixed state.

\begin{theorem}[one-shot generalized equality]
\label{thm:GEQimproved-revised}
For every \(d\geq1\),
\[
 Q^{\mathrm{PSQM},*}(\operatorname{GEQ}_{k,d})
 \leq
 \begin{cases}
 k\lceil d/2\rceil,&k\text{ even},\\
 2k\lceil d/4\rceil,&k\text{ odd}.
 \end{cases}
\]
In particular, if \(2\mid d\) for even \(k\), or \(4\mid d\) for odd \(k\), then
\[
 Q^{\mathrm{PSQM},*}(\operatorname{GEQ}_{k,d})\leq kd/2.
\]
These are one-shot bounds for one length-\(d\) instance.
\end{theorem}

\begin{proof}
For even \(k\), group coordinates into pairs and apply the $Sum_2$ protocol of \cite{KN21} to each pair.  
For odd \(k\), group coordinates into blocks of four and apply Lemma~\ref{lem:oddsecure} to each block, 
padding by at most three zero coordinates. This gives the aggregate-dependent family \((\Omega_s)_s\).  

The players then apply the random input transformation by $\lambda$ defined above 
and produce the family $(\overline{\Omega}_s)_s$ defined in \eqref{eq:density-twirl-revised}.  
Correctness follows from the correctness of the secure-sum layer and the fact that \(s_\lambda=0\) if and only if \(s=0\). 
Perfect privacy follows from \eqref{eq:zero-nonzero-states-revised}.
\end{proof}

The coordinate grouping occurs inside one vector-valued input.  It does
not invoke multiple copies of generalized equality and hence gives an ordinary one-shot communication bound.

\subsection{Our upper bound}\label{subsec:all-EQ-main}

We now combine the local embedding with the generalized-equality protocol to obtain the claimed upper bound for $\AEQ_{k,n}$.

\begin{theorem} \label{thm:AEQimproved-revised}
For every fixed \(k\geq3\),
\[
 Q^{\mathrm{PSQM},*}(\AEQ_{k,n})
 \leq \frac{k}{2h_2(1/k)}n+O_k(\sqrt n).
\]
There is no asymptotic parity distinction.
\end{theorem}

\begin{proof}
Apply Lemma~\ref{lem:AEQkembed-revised}, and then apply Theorem~\ref{thm:GEQimproved-revised} to the resulting length-\(d\) generalized-equality instance. 
For even \(k\), the communication cost is \[ k\left\lceil\frac d2\right\rceil = \frac{kd}{2}+O_k(1), \] whereas for odd \(k\), it is 
\[
 2k\left\lceil\frac d4\right\rceil = \frac{kd}{2}+O_k(1). 
\] 
Since $d = \frac{n}{h_2(1/k)} + O_k(\sqrt n)$, both cases give $\frac{k}{2h_2(1/k)}n + O_k(\sqrt n)$. 

Correctness follows from exact preservation of the zero-sum condition.
For perfect privacy, apply the random input transformation defined above. 
Equations \eqref{eq:density-twirl-revised} and \eqref{eq:zero-nonzero-states-revised} show that 
the joint state received by the referee depends only on whether the embedded aggregate is zero, 
which is equivalent to all original inputs being equal.
\end{proof}

\begin{corollary}
\[
 Q^{\mathrm{PSQM},*}(\AEQ_{3,n}) \leq\frac{3}{2h_2(1/3)}n+O(\sqrt n) = (1.633\ldots) n + O(\sqrt n).
\]
\end{corollary}

\subsection{Shared-randomness-only comparison}

The following lower bound follows readily from perfect correctness, but we include the proof for completeness.

\begin{proposition} 
Every perfectly correct PSQM protocol for \(\AEQ_{k,n}\) that uses shared classical randomness but no prior entanglement communicates at least \(kn\) qubits. 
\end{proposition} 

\begin{proof} 
Fix a player \(j\) and distinct inputs \(x,x'\in\{0,1\}^n\), and set every input other than that of player \(j\) equal to \(x\). 
The all-\(x\) input $(x,x,\ldots,x)$ has output \(1\), whereas replacing player \(j\)'s input by \(x'\) gives output \(0\). 
Since every conditional success probability is at most one, an average success probability of one implies perfect correctness 
for every \(r\in\mathcal R\). 
Moreover, for every such \(r\), the message state conditioned on \(r\) is a positive component of the corresponding averaged message state. 
Since the support of every positive component is contained in the support of the averaged state, 
the conditional message states corresponding to the two different outputs have orthogonal supports. 
Fix \(r\in\mathcal R\). 
Since the protocol uses no prior entanglement, the initial shared state has the product form 
\[
 \sigma_{\mathcal A_1\cdots \mathcal A_k} = \bigotimes_{i=1}^k \sigma_i. 
\] 
For each player \(i\), define 
\[
 \rho_i(x;r) := \mathcal E_i^{x,r}(\sigma_i). 
\] 
The two global message states conditioned on \(r\) are then 
\[
 \bigotimes_{i=1}^k \rho_i(x;r) \qquad\text{and}\qquad \rho_j(x';r) \otimes \bigotimes_{i\ne j}\rho_i(x;r), 
\] respectively. 
Their Hilbert--Schmidt overlap therefore factors as 
\[
 \operatorname{Tr}\!\left( \rho_j(x;r)\rho_j(x';r) \right) \prod_{i\ne j} \operatorname{Tr}\!\left( \rho_i(x;r)^2 \right) = 0. 
\] 
Every factor in the product over \(i\ne j\) is strictly positive, because each \(\rho_i(x;r)\) is a density operator. 
Hence \[ \operatorname{Tr}\!\left( \rho_j(x;r)\rho_j(x';r) \right) = 0. \] Since the two operators are positive semidefinite, their supports are orthogonal. 
The same argument applies to every pair of distinct \(x,x'\in\{0,1\}^n\). Thus, for the fixed \(r\), the \(2^n\) local message states of player \(j\) have pairwise orthogonal nonzero supports. 
Therefore, player \(j\) must communicate at least $n$ qubits. Since \(j\) was arbitrary, summing over all \(k\) players gives a total communication cost of at least \(kn\) qubits. 
\end{proof}

The coefficient in the entanglement-assisted upper bound of Theorem~\ref{thm:AEQimproved-revised}, $\frac{k}{2h_2(1/k)}$, is smaller than $k$ for $3\leq k\leq 9$.   
Thus the entanglement-assisted upper bound is strictly smaller than the shared-randomness-only lower bound \(kn\) throughout this range. 


\section{Three-Bit Boolean Functions}

In this section, we investigate PSQM protocols for three-bit Boolean functions using one shared GHZ state and one transmitted qubit per player. 
We first give explicit protocols for majority and equality, 
and then extend these constructions to an analytic family of \(80\) functions in the randomized local-Pauli model. 
Finally, we compare this analytic family with an exhaustive linear-programming search within the same restricted model. 

Throughout this section, Pauli operators are identified up to global phase, and we use the representatives $\{I,X,Z,XZ\}$. 
In particular, $Y=iXZ$ is represented by $XZ$. 

\subsection{GHZ labels}

For the three-bit constructions, let
\[
 \ket{G_+}=\frac{\ket{000}+\ket{111}}{\sqrt2}
\]
and define an orthonormal basis by
\[
 \ket{\Phi_{u,v,w}} =(I\otimes X^u\otimes X^vZ^w)\ket{G_+},
 \qquad u,v,w\in\F_2.
\]
Throughout this section, whenever we refer to the three-qubit \emph{GHZ basis}, we mean the basis defined above. 
We call $(u,v,w)$ the GHZ label of $\ket{\Phi_{u,v,w}}$. 

\begin{lemma}[Pauli-to-GHZ-label rule]\label{lem:GHZlabel}
If player 1, player 2, and player 3 apply $X^{a_1}Z^{b_1}$, $X^{a_2}Z^{b_2}$, and $X^{a_3}Z^{b_3}$, respectively, 
the resulting state is, up to global phase, $\ket{\Phi_{u,v,w}}$ with
\[
 u=a_1\oplus a_2,\qquad
 v=a_1\oplus a_3,\qquad
 w=b_1\oplus b_2\oplus b_3.
\]
\end{lemma}

\begin{proof}
The stabilizer identity $(X\otimes X\otimes X)\ket{G_+}=\ket{G_+}$ reduces the bit-flip information to two relative parities. 
A Pauli-$Z$ operation on any one qubit toggles the relative phase between $\ket{000}$ and $\ket{111}$, 
and thus only the parity of the three $Z$ exponents remains.
\end{proof}

\subsection{Majority and equality} \label{subsec:three-bit-majority} 

We first derive the three-qubit protocol for the majority function $\MAJ(x,y,z) := \mathbf{1}[x+y+z\ge 2]$ 
from the optimal classical PSM protocol of Shinagawa and Nuida~\cite{ShinagawaNuida2023}. 
This derivation explains both the choice of local Pauli encodings and the two four-dimensional subspaces associated with the two output values. 
We then compress the shared randomness to four values while preserving the same two output-dependent density operators.

\paragraph{The classical \texorpdfstring{$(4,4,4)$}{(4,4,4)} protocol.} 

We first recall the classical \((4,4,4)\) PSM protocol of Shinagawa and Nuida~\cite{ShinagawaNuida2023},  
where each of the three players sends a message from a four-element alphabet, and hence the total communication cost is \(6\) bits.

The players share independent uniform random variables $b\in\{0,1\}$ and $c_1,c_2\in\mathbb{Z}_4$, and set $c_3=-c_1-c_2 \pmod 4$.  
For \(b=0\), the players send 
\[ 
m_i=x_i+c_i \pmod 4, \qquad i\in\{1,2,3\}. 
\] 
For \(b=1\), they send 
\[ 
m_1=1-x_1+c_1,\qquad m_2=-x_2+c_2,\qquad m_3=-x_3+c_3 \pmod 4. 
\] 
The referee computes 
\[ 
m=m_1+m_2+m_3\pmod 4 
\] 
and outputs \(0\) if \(m\in\{0,1\}\), and \(1\) if \(m\in\{2,3\}\). 

For \(b=0\), the sum is 
\[ m=x_1+x_2+x_3\pmod 4, \] whereas for \(b=1\), \[ m=1-(x_1+x_2+x_3)\pmod 4. \] 
Consequently, if \(x_1+x_2+x_3\in\{0,1\}\), then \(m\) is uniform on \(\{0,1\}\), 
while if \(x_1+x_2+x_3\in\{2,3\}\), then \(m\) is uniform on \(\{2,3\}\). 
For each fixed value of \(m\), 
the uniform choices of \(c_1\) and \(c_2\), with \(c_3=-c_1-c_2\), make the joint message tuple uniform over all triples in \(\mathbb{Z}_4^3\) 
whose coordinate sum is \(m\). Since \(b\) makes \(m\) uniform over \(\{0,1\}\) or over \(\{2,3\}\) according to the majority value, 
the global message distribution depends only on that value.
Hence the protocol is perfectly correct and perfectly private, 
with message alphabet \(\mathbb{Z}_4\) for each player. 

\paragraph{A Pauli lift of the classical protocol.}

We now convert the classical protocol into a quantum protocol by replacing each classical message value with a fixed local Pauli operation. 
The shared randomness and the local message functions of the classical protocol remain unchanged. 

For each \(m\in\mathbb Z_4\), define \[ U_m := X^{a(m)}Z^{d(m)}, \] where 
\begin{equation}\label{eq:pauli-ad} 
(a(0),d(0))=(0,0),\quad (a(1),d(1))=(1,1),\quad (a(2),d(2))=(0,1),\quad (a(3),d(3))=(1,0). 
\end{equation} 
Thus, up to global phase, \[ U_0=I,\qquad U_1=XZ,\qquad U_2=Z,\qquad U_3=X. \] 

Suppose that the three players share one copy of $\ket{G^+} = \frac{\ket{000}+\ket{111}}{\sqrt2}$. 
For a global input $x=(x_1,x_2,x_3)\in\{0,1\}^3$ and shared randomness $r=(b,c_1,c_2) \in \{0,1\}\times\mathbb Z_4^2$,  
set $c_3=-c_1-c_2\pmod 4$, and let $m_i(x_i;r)\in\mathbb Z_4$ denote the message sent by player \(i\) in the classical protocol above. 
The corresponding conditional quantum message state is 
\[ \ket{\psi_{x;r}} := \left( U_{m_1(x_1;r)} \otimes U_{m_2(x_2;r)} \otimes U_{m_3(x_3;r)} \right) \ket{G^+}. \] 

The GHZ label of this state is a deterministic function of the classical joint message tuple. 
More precisely, define $\Lambda: \mathbb Z_4^3 \longrightarrow \mathbb F_2^3$ by 
\[
 \Lambda(m_1,m_2,m_3) := \left( a(m_1)\oplus a(m_2), a(m_1)\oplus a(m_3), d(m_1)\oplus d(m_2)\oplus d(m_3) \right). 
\] 
By Lemma~\ref{lem:GHZlabel}, 
the state \(\ket{\psi_{x;r}}\) is, up to global phase, 
the GHZ-basis state indexed by 
\[
 \Lambda\bigl( m_1(x_1;r), m_2(x_2;r), m_3(x_3;r) \bigr). 
\] 

We next relate this GHZ label to the quantity decoded by the classical referee. 
Write \[ m_i=2q_i+p_i, \qquad p_i,q_i\in\mathbb F_2. \] 
Under the correspondence (\ref{eq:pauli-ad}) above, 
\[ a(m_i)=p_i, \qquad d(m_i)=p_i\oplus q_i. \] 
Hence, if 
\[
 (u,v,w)=\Lambda(m_1,m_2,m_3), 
\] then \[ u=p_1\oplus p_2, \qquad v=p_1\oplus p_3. \] 

Now write 
\[
 m_1+m_2+m_3=2q+p\pmod 4, \qquad p,q\in\mathbb F_2. 
\] 
Binary addition, including the carry from the three low-order bits, gives 
\[
 p=p_1\oplus p_2\oplus p_3 
\] and 
\[
 q = q_1\oplus q_2\oplus q_3 \oplus p_1p_2 \oplus p_1p_3 \oplus p_2p_3. 
\] 
Using $p_2=p_1\oplus u$ and $p_3=p_1\oplus v$, we obtain 
\begin{align*}
 w &= (p_1\oplus q_1)\oplus (p_2\oplus q_2)\oplus (p_3\oplus q_3)\\ 
   &= q\oplus p_1\oplus p_2\oplus p_3\oplus p_1p_2 \oplus p_1p_3\oplus p_2p_3\\
   &= q\oplus u\oplus v\oplus uv. 
\end{align*} 
Therefore, 
\[ w= \begin{cases} u\oplus v\oplus uv, &m_1+m_2+m_3\in\{0,1\}\pmod 4,\\ 1\oplus u\oplus v\oplus uv, &m_1+m_2+m_3\in\{2,3\}\pmod 4. \end{cases} 
\]

The classical referee outputs \(0\) in the first case and \(1\) in the second. Thus the GHZ labels corresponding to the two output values lie in 
\[ S_0 := \left\{ (u,v,u\oplus v\oplus uv): u,v\in\mathbb F_2 \right\} = \{000,011,101,111\} \] and 
\[ S_1 := \left\{ (u,v,1\oplus u\oplus v\oplus uv): u,v\in\mathbb F_2 \right\} = \{001,010,100,110\}, \] respectively. 
Since $S_0\cap S_1=\varnothing$, 
a GHZ-basis measurement followed by membership testing in \(S_1\) recovers the majority value with certainty. 
Hence the quantum protocol is perfectly correct.

Perfect privacy follows directly from the classical protocol. 
The distribution of the classical joint message tuple $(m_1,m_2,m_3)$ depends on the input only through its majority value. 
Since the GHZ label is its deterministic image under \(\Lambda\), the induced GHZ-label distribution, 
and hence the joint quantum state received by the referee, also depends only on the majority value.

We have proved the following theorem. 

\begin{theorem}[Pauli lift of the classical majority protocol] \label{thm:maj-direct-quantization}
The classical \((4,4,4)\)-PSM protocol for \(\operatorname{MAJ}_3\) in~\cite{ShinagawaNuida2023} can be converted into 
a perfectly correct and perfectly private PSQM protocol by replacing each classical message \(m\in\mathbb Z_4\) 
with the local Pauli operation \(U_m\) defined above. 
The converted protocol uses one shared three-qubit GHZ state, five shared random bits, and one transmitted qubit per player. 
\end{theorem} 

We next determine the two output-dependent density operators $\varrho_0$ and $\varrho_1$, 
the joint state received by the referee when the output is $0$ and $1$, respectively. 
For each fixed value of $m_1+m_2+m_3\pmod 4$,  
each pair \((u,v)\in\mathbb F_2^2\) has exactly four preimages among the classical transcripts. 
Indeed, for fixed $m=2q+p$ and fixed $(u,v)$, the equations above determine $p_1,p_2,p_3$ uniquely, 
while the high-order bits $q_1,q_2,q_3$ satisfy one affine constraint and therefore admit exactly four choices. 
Moreover, for each majority value, the classical protocol makes the corresponding two possible values of \(m_1+m_2+m_3\) equally likely. 
Hence every GHZ label in \(S_0\) or \(S_1\), as appropriate, occurs with probability \(1/4\). Therefore, 
\begin{equation} \label{eq:4.4}
\varrho_0 = \frac14 \sum_{\lambda\in S_0} \ket{\Phi_\lambda}\!\bra{\Phi_\lambda}, 
\end{equation} 
whereas 
\begin{equation} \label{eq:4.5}
 \varrho_1 = \frac14 \sum_{\lambda\in S_1} \ket{\Phi_\lambda}\!\bra{\Phi_\lambda}. 
\end{equation}

It is worth emphasizing that the correspondence $0,1,2,3\longmapsto I,XZ,Z,X$ is not a group homomorphism from \(\mathbb Z_4\) to the one-qubit Pauli group modulo phase. The construction therefore does not coherently implement addition in \(\mathbb Z_4\). 
Rather, it maps the classical transcript distribution to a coarser GHZ-label distribution. 
Perfect privacy is preserved under this deterministic postprocessing, 
while the compatibility between the classical masking and the GHZ stabilizer structure maps the two majority values to the disjoint GHZ-label sets \(S_0\) and \(S_1\).

\paragraph{Observable formulation.} 
The two output states also admit a compact operator description. Define 
\[
M := X\otimes X\otimes X +X\otimes Y\otimes Y +Y\otimes X\otimes Y +Y\otimes Y\otimes X. 
\]
The eigenvalues of \(M\) are \(+2\) and \(-2\), each with multiplicity four, and $\varrho_0-\varrho_1=-\frac{1}{8}M$. Equivalently, 
\[ 
\varrho_0 = \frac{1}{8}\left(I-\frac{M}{2}\right), \qquad \varrho_1 = \frac{1}{8}\left(I+\frac{M}{2}\right). 
\] 
Hence the referee may decode the majority value by measuring the sign of the observable \(M\). 
The \(-2\) eigenspace corresponds to output \(0\), and the \(+2\) eigenspace corresponds to output \(1\). 

\paragraph{Shared-randomness compression.} 
In the direct quantization, the intermediate classical messages \(m_1,m_2,m_3\), and hence the local Pauli operations, 
are determined by $(b,c_1,c_2)\in\{0,1\}\times\mathbb Z_4^2$. 
The shared-randomness support therefore has size \(32\). 
We now show that the same two output-dependent states can be realized using only four equally likely local-Pauli encoding types, 
without the \(\mathbb Z_4\)-valued masks.

The four encoding types admit the following direct construction. 
Choose \((s,t)\in\mathbb F_2^2\) uniformly at random and set 
\[
 u=x\oplus y\oplus s, \qquad v=x\oplus z\oplus t. 
\] 
For every fixed input, the pair \((u,v)\) is then uniform on \(\mathbb F_2^2\). 
To ensure that the resulting GHZ label belongs to \(S_{\operatorname{MAJ}_3(x,y,z)}\), 
we should set 
\[
 w = \operatorname{MAJ}_3(x,y,z) \oplus u\oplus v\oplus uv. 
\] 
Using \(\operatorname{MAJ}_3=xy\oplus xz\oplus yz\), this simplifies to 
\[
 w = (1\oplus s\oplus t)x \oplus(1\oplus t)y \oplus(1\oplus s)z \oplus s\oplus t\oplus st, 
\] 
which is affine in the three inputs for every fixed \((s,t)\). 

To implement these affine GHZ-label maps, 
choose the local Pauli exponents for players $1,2,3$, respectively, as 
\[
 \begin{aligned} 
a_1&=x, & b_1&=(1\oplus s\oplus t)x,\\ 
a_2&=y\oplus s, & b_2&=(1\oplus t)y,\\ 
a_3&=z\oplus t, & b_3&=(1\oplus s)z\oplus s\oplus t\oplus st,  
\end{aligned} 
\] 
so that the corresponding local Pauli operations for players $1,2,3$ are, respectively,  
\[
 \begin{aligned} 
A_x^{s,t} &:= X^xZ^{(1\oplus s\oplus t)x},\\ 
B_y^{s,t} &:= X^{y\oplus s}Z^{(1\oplus t)y},\\ 
C_z^{s,t} &:= X^{z\oplus t} Z^{(1\oplus s)z\oplus s\oplus t\oplus st}. 
\end{aligned} 
\] 
Then Lemma~\ref{lem:GHZlabel} gives 
\[
 u=a_1\oplus a_2=x\oplus y\oplus s, \qquad v=a_1\oplus a_3=x\oplus z\oplus t, 
\] and 
\[
 w=b_1\oplus b_2\oplus b_3=\operatorname{MAJ}_3(x,y,z) \oplus u\oplus v\oplus uv, 
\] 
as desired. 
Evaluating these expressions for the four choices of \((s,t)\in\mathbb F_2^2\) gives the following array, 
in which the superscripts $s,t$ are omitted from the column headings.

\[
\begin{array}{c|cc|cc|cc} 
(s,t) & A_0 & A_1 & B_0 & B_1 & C_0 & C_1 \\ \hline
 (0,0) & I & XZ & I & XZ & I & XZ \\
 (0,1) & I & X & I & X & XZ & I \\
 (1,0) & I & X & X & Z & Z & XZ \\
 (1,1) & I & XZ & X & I & XZ & Z 
\end{array}
\] 
By Lemma~\ref{lem:GHZlabel}, these encodings generate the GHZ labels shown in Table~\ref{tab:maj-four}. 

\begin{table}[htbp] \centering  
\begin{tabular}{c|cccc|c} \hline 
$xyz$ & $(s,t)=(0,0)$ & $(s,t)=(0,1)$ & $(s,t)=(1,0)$ & $(s,t)=(1,1)$ & $\MAJ(x,y,z)$ \\ \hline 
$000$ & $(0,0,0)$ &$(0,1,1)$ & $(1,0,1)$ &  $(1,1,1)$ & $0$ \\ 
$001$ & $(0,1,1)$ &$(0,0,0)$ & $(1,1,1)$ &  $(1,0,1)$ & $0$ \\ 
$010$ & $(1,0,1)$ &$(1,1,1)$ & $(0,0,0)$ &  $(0,1,1)$ & $0$ \\ 
$100$ & $(1,1,1)$ &$(1,0,1)$ & $(0,1,1)$ &  $(0,0,0)$ & $0$ \\ 
$011$ & $(1,1,0)$ &$(1,0,0)$ & $(0,1,0)$ &  $(0,0,1)$ & $1$ \\ 
$101$ & $(1,0,0)$ &$(1,1,0)$ & $(0,0,1)$ &  $(0,1,0)$ & $1$ \\ 
$110$ & $(0,1,0)$ &$(0,0,1)$ & $(1,1,0)$ &  $(1,0,0)$ & $1$ \\ 
$111$ & $(0,0,1)$ &$(0,1,0)$ & $(1,0,0)$ &  $(1,1,0)$ & $1$ \\ \hline 
\end{tabular} 
\caption{GHZ labels generated by the compressed majority protocol}\label{tab:maj-four}
\end{table}

Every row corresponding to output \(0\) is a permutation of \(S_0\), whereas every row corresponding to output \(1\) is a permutation of \(S_1\). 
Consequently, the protocol with four equally likely encoding types produces exactly the same two density operators \(\varrho_0\) and \(\varrho_1\) 
as the direct quantization with shared-randomness support of size \(32\).

We have proved the following theorem. 

\begin{theorem}[Compressed three-bit majority protocol] \label{thm:three-bit-majority} 
There is a perfectly correct and perfectly private PSQM protocol for \(\MAJ\) using one shared three-qubit GHZ state, 
two shared random bits, local Pauli encodings, and one transmitted qubit per player. 
Its output-dependent states are the states \(\varrho_0\) and \(\varrho_1\) in~\eqref{eq:4.4} and~\eqref{eq:4.5}. 
\end{theorem} 


The protocol using four equally likely encoding types should therefore be viewed as a shared-randomness-compressed realization of the direct Pauli quantization of the classical \((4,4,4)\) protocol. The classical protocol transmits the complete \(\mathbb Z_4\)-valued message tuple, whereas the quantum protocol maps this tuple to a coarser GHZ-label distribution that retains only the information needed to determine the function value.

\paragraph{Equality by a parity twist.} 
Let $\PAR(x,y,z)=x\oplus y\oplus z$ and $\EQ(x,y,z)=\mathbf{1}[x=y=z]$. Their truth tables satisfy 
\[ \EQ = 1\oplus\MAJ\oplus\PAR. \] 
Moreover, 
\[
 (Z^x\otimes Z^y\otimes Z^z) \ket{\Phi_{u,v,w}} = \ket{\Phi_{u,v,w\oplus x\oplus y\oplus z}} 
\] up to global phase. 
Thus the third GHZ-label component is flipped exactly when \(\operatorname{PAR}_3(x,y,z)=1\). 
Since flipping this component exchanges \(S_0\) and \(S_1\), we obtain the following consequence. 

\begin{corollary}[Three-bit equality] \label{cor:eq3}
Composing the encoders in Theorem~\ref{thm:three-bit-majority} with the local operations \(Z^x,Z^y,Z^z\), 
and reversing the referee's output, gives a perfectly correct and perfectly private protocol for \(\EQ\) with the same resources. 
\end{corollary}

In~\cite{ShinagawaNuida2023}, the authors showed that the classical PSM complexity of $\EQ$ is $3\log_2 3$. 
Thus the above corollary shows quantum communication advantage for $\EQ$. 

\subsection{An analytic family of 80 functions}

Every three-bit Boolean function has a unique algebraic normal form
\[
 f=c\oplus ax\oplus by\oplus dz \oplus\alpha xy\oplus\beta xz\oplus\gamma yz\oplus\delta xyz.
\]

\begin{theorem}[$80$-function construction]\label{thm:80}
Suppose 
\[
\delta=0\quad \mbox{and} \quad (\alpha,\beta,\gamma) \in\{(0,0,0),(1,0,0),(0,1,0),(0,0,1),(1,1,1)\}.
\] 
Then $f$ admits a perfectly correct and perfectly private protocol with one shared three-qubit GHZ state, finite shared randomness, local Pauli encodings, and one transmitted qubit per player.  
These conditions describe exactly $5\cdot2^4=80$ distinct functions.
\end{theorem}

\begin{proof}
Under permutations of the players, it suffices to consider the three quadratic coefficient vectors 
\[ (0,0,0),\qquad (1,0,0),\qquad (1,1,1). \] 
These correspond, respectively, to an affine function, a function with one quadratic term such as \(xy\), 
and a function containing all three quadratic terms \[ xy\oplus xz\oplus yz = \operatorname{MAJ}_3. \] 
We treat these three cases using affine phase encodings, the \(xy\oplus z\) base protocol, and the majority protocol, respectively. 
The base protocol also yields a protocol for $xy$ by having the third player ignore its input and always use its $z=0$ encoder. 
Player permutations cover the single-term cases \(xz\) and \(yz\). 
We also use input complementation, which swaps a player's two encoders, and output complementation, which reverses the referee's declaration.

All 16 affine functions are immediate: for case $(0,0,0)$; 
\[
 \ell=c\oplus ax\oplus by\oplus dz,
\]
the players apply $Z^{ax},Z^{by},Z^{dz}$ and the referee reads the GHZ phase, reversing the result if $c=1$.

Next, we consider case $(1,0,0)$. 
For the base function \(xy\oplus z\), let the shared randomness be uniform on $\mathbb F_2^2$. 
The corresponding local-Pauli encodings are given in the following array. 
\[
\begin{array}{c|cc|cc|cc}
 (s,t)&A_0&A_1&B_0&B_1&C_0&C_1\\ \hline
(0,0)&I&XZ&I&XZ&I&Z\\
(0,1)&I&X&I&X&XZ&X\\
(1,0)&I&X&X&Z&Z&I\\
(1,1)&I&XZ&X&I&XZ&X
\end{array}
\]
The GHZ-label table is as follows. 
\[
\begin{array}{c|cccc|c}
xyz  &(0,0)&(0,1)&(1,0)&(1,1)&xy\oplus z\\ \hline
000  &(0,0,0)&(0,1,1)&(1,0,1)&(1,1,1)&0\\
001  &(0,0,1)&(0,1,0)&(1,0,0)&(1,1,0)&1\\
010  &(1,0,1)&(1,1,1)&(0,0,0)&(0,1,1)&0\\
011  &(1,0,0)&(1,1,0)&(0,0,1)&(0,1,0)&1\\
100  &(1,1,1)&(1,0,1)&(0,1,1)&(0,0,0)&0\\
101  &(1,1,0)&(1,0,0)&(0,1,0)&(0,0,1)&1\\
110  &(0,1,0)&(0,0,1)&(1,1,0)&(1,0,0)&1\\
111  &(0,1,1)&(0,0,0)&(1,1,1)&(1,0,1)&0
\end{array}
\]
Every input $xy\oplus z=0$ produces the uniform distribution on $S_0$, 
whereas every input with $xy\oplus z=1$ produces the uniform distribution on $S_1$. 

This protocol also gives a protocol for $xy$: 
player $3$ ignores its input and always applies the encoder $C_0$.\footnote{Although \(xy\) can be computed by applying an existing two-party protocol for \(\operatorname{AND}_2\) to the first two inputs and ignoring the third input, 
we require here a realization within the fixed three-party GHZ--Pauli model. 
} Equivalently, we restrict the $xy\oplus z$ protocol to $z=0$. 
The resulting protocol produces the uniform distribution on $S_0$ when $xy=0$ and the uniform distribution on $S_1$ when $xy=1$.

Now fix
\[
 f=xy\oplus ax\oplus by\oplus dz\oplus c.
\]
If $d=0$, use
\[
 (x\oplus b)(y\oplus a)=xy\oplus ax\oplus by\oplus ab
\]
and obtain $f$ from the $xy$ protocol by input and output complementations.  
If $d=1$, use
\[
 (x\oplus b)(y\oplus a)\oplus z
 =xy\oplus ax\oplus by\oplus z\oplus ab
\]
and the $xy\oplus z$ protocol.  
Permuting players covers $xz$ and $yz$.  This yields $3\cdot16=48$ functions.

Finally, we consider case $(1,1,1)$; $\MAJ=xy\oplus xz\oplus yz$. 
For the majority protocol, 
composing the local encoders with \(Z^{ax}\), \(Z^{by}\), and \(Z^{dz}\), respectively, exchanges the two output-dependent states whenever 
\[
 ax\oplus by\oplus dz=1. 
\] 
Output complementation then supplies the constant term \(c\). 
This gives 16 functions whose quadratic part is $xy\oplus xz\oplus yz$. 
Together with the \(16\) affine functions and the \(48\) functions whose quadratic part consists of exactly one of $xy,xz,yz$, this proves Theorem~\ref{thm:80}.  
\end{proof}

\begin{remark}
The arrays of $\MAJ$ and $xy\oplus z$ differ only in the \(C_1\) column. 
More precisely, the \(xy\oplus z\) array is obtained from the majority array by composing \(C_1\) with 
\[ X,\quad X,\quad XZ,\quad XZ \] for \((s,t)=(0,0),(0,1),(1,0),(1,1)\), respectively. 
To see why this works, define 
\[
 D(u,v,w):=w\oplus u\oplus v\oplus uv, 
\] 
so that \(D(u,v,w)=b\) exactly when \((u,v,w)\in S_b\). 
Composing the third player's operation with \(X\) changes \(D\) by \(1\oplus u\), 
while composing it with \(XZ\) changes \(D\) by \(u\). 
In the majority array, 
\[
 u= \begin{cases} x\oplus y,&(s,t)=(0,0),(0,1),\\ 1\oplus x\oplus y,&(s,t)=(1,0),(1,1). \end{cases} 
\] 
Hence all four corrections change the decoded bit by \(1\oplus x\oplus y\) when \(z=1\), and by zero when \(z=0\). 
Since \[
 \operatorname{MAJ}_3(x,y,z) \oplus(xy\oplus z) = z(1\oplus x\oplus y), 
\] 
the corrected array computes \(xy\oplus z\).
\end{remark}

\subsection{Computational exploration of a restricted GHZ--Pauli model}
\label{sec:computational-exploration}

The analytic constructions above were motivated in part by a finite computational exploration. This exploration is restricted to the model described below, 
and its negative outcomes are reported only as numerical observations.

\paragraph{Restricted model.}
We restrict attention to protocols satisfying all of the following conditions.
\begin{enumerate}
  \item The shared quantum resource is the fixed GHZ state $\ket{G_+}=\frac{\ket{000}+\ket{111}}{\sqrt 2}$.
  \item The players may use an arbitrary input-independent classical random variable \(r\in\mathcal R\), hidden from the referee.  No a priori bound is imposed on its finite support size. 
  \item For every \(r\in\mathcal R\), every player \(i\in\{1,2,3\}\), and every input bit \(b\in\{0,1\}\), 
the local encoding operation is given by a Pauli operator in $\{I,X,Z,XZ\}$, up to global phase.
  \item Each player sends its share of the GHZ state to the referee.
  \item The referee may perform an arbitrary measurement.
\end{enumerate}
Thus the search is complete only within the one-GHZ randomized local-Pauli model. 
It does not cover arbitrary local unitary encodings or, more generally, arbitrary local CPTP encodings, including encodings that couple the message qubits to auxiliary systems. Nor does it cover shared states other than the fixed GHZ state, multiple shared entangled states, or more than one transmitted qubit per player.

Every deterministic shared-randomness value is specified by
\[
 t=(A_0,A_1;B_0,B_1;C_0,C_1) \in\{I,X,Z,XZ\}^6.
\]
Hence there are exactly $4^6=4096$ deterministic encoding types. 
By the GHZ-label rule, a type $t$ induces a function $L_t:\{0,1\}^3\longrightarrow\mathbb F_2^3$, where $L_t(u)$ is the GHZ label of the state received on input $u$. 
Different shared-randomness values may induce the same deterministic encoding type. 
Combining their probabilities, any protocol in the restricted class is described by a probability distribution \((p_t)_t\) over the \(4096\) types.

For an input $u\in\{0,1\}^3$, define the induced GHZ-label distribution by
\begin{equation}
 D_u(\lambda)
 :=\sum_t p_t\,\mathbf 1[L_t(u)=\lambda],
 \qquad \lambda\in\mathbb F_2^3.
 \label{eq:label-distribution-computation}
\end{equation}
The global message state is
\[
 \rho_u
 =\sum_{\lambda\in\mathbb F_2^3}
 D_u(\lambda)\proj{\Phi_\lambda}.
\]
In particular, all averaged message states in this restricted model are diagonal in the same GHZ basis.

Fix a Boolean function $f:\{0,1\}^3\longrightarrow\{0,1\}$ and a subset $S\subseteq\mathbb F_2^3$ interpreted as the set of GHZ labels decoded as one.  
We retain only those deterministic types that satisfy
\begin{equation}
 L_t(u)\in S
 \quad\Longleftrightarrow\quad
 f(u)=1
 \qquad\text{for every }u\in\{0,1\}^3.
 \label{eq:correct-types}
\end{equation}
Let $\mathcal T_{f,S}$ be the set of retained types. 
For this fixed pair $(f,S)$, we solve the following linear feasibility problem in the variables $p_t$, $t\in\mathcal T_{f,S}$:
\begin{align}
 &p_t\geq 0\ \ \ \ (t\in\mathcal T_{f,S}), \label{eq:lp-nonnegative}\\
 &\sum_{t\in\mathcal T_{f,S}}p_t =1, \label{eq:lp-normalization}\\
 &\sum_{t\in\mathcal T_{f,S}}p_t \bigl( \mathbf 1[L_t(u)=\lambda] -\mathbf 1[L_t(v)=\lambda] \bigr)=0, \label{eq:lp-privacy}
\end{align}
for every $u,v$ satisfying $f(u)=f(v)$ and every $\lambda\in\mathbb F_2^3$. 
Equation~\eqref{eq:lp-privacy} is exactly the perfect-privacy requirement \(D_u=D_v\) whenever \(f(u)=f(v)\).
Equation~\eqref{eq:correct-types} ensures perfect correctness for every deterministic type receiving positive probability.

It is sufficient to enumerate all \(2^8=256\) subsets \(S\subseteq\mathbb F_2^3\). 
Indeed, all received states in the restricted model are diagonal in the GHZ basis. 
If the two output-dependent states have orthogonal supports, 
then their GHZ-label supports are disjoint and can be separated by some subset \(S\subseteq\mathbb F_2^3\). 
Consequently, an arbitrary perfectly correct referee measurement can be replaced, in this restricted model, 
by a GHZ-basis measurement followed by membership testing in $S$.

The computation enumerated all \(2^8=256\) Boolean truth tables, all \(256\) decoder supports, and all \(4096\) deterministic encoding types. 
The constraint matrices of the resulting feasibility problems have entries in \(\{-1,0,1\}\) 
since each privacy constraint is the difference of two indicator sums. 
These problems were solved in floating-point arithmetic using the HiGHS linear-programming solver through \texttt{scipy.optimize.linprog}~\cite{HuangfuHall2018,VirtanenEtAl2020}. 
For every reported feasible instance, the returned probability distribution was independently checked 
by recomputing all eight distributions in~\eqref{eq:label-distribution-computation} and verifying 
\[ f(u)=f(v) \quad\Longrightarrow\quad D_u=D_v. \] 
We also checked normalization, nonnegativity, and disjointness of the GHZ label supports corresponding to the two function values.

\begin{remark}[Computational outcome and its interpretation] \label{rem:computational-outcome}
The computation found feasible solutions for \(80\) of the \(256\) three-bit Boolean functions. 
Classified by the number of inputs on which \(f\) takes the value \(1\), 
the counts were as follows.
\[ 
\begin{array}{c|ccccc} |f^{-1}(1)|&0&2&4&6&8\\ \hline \text{number found feasible}&1&16&46&16&1 
\end{array} 
\]
The positive cases are no longer dependent on the computation:
Theorem~\ref{thm:80} gives explicit analytic constructions for all $80$ functions.

For the other functions, the floating-point solver reported no feasible solution in the restricted model. 
This group includes $\operatorname{EXACT}_1$, $\operatorname{EXACT}_2$, where $\operatorname{EXACT}_t(x,y,z):=\mathbf1[x+y+z=t]$ and the sum is taken over the integers,  
as well as the $48$ affine extensions of the two-edge quadratic forms with coefficient vectors $(1,1,0),(1,0,1),(0,1,1)$. 
We record these negative outcomes only as numerical observations, not as rigorous impossibility results.
In particular, they imply no impossibility statement for arbitrary local \(U(2)\) encodings, general CPTP encodings, other shared states, or larger quantum messages. 
\end{remark}

The purpose of the computation is therefore exploratory. 
It identifies the algebraic pattern captured by Theorem~\ref{thm:80}, 
suggests that a two-edge quadratic form such as $xy\oplus xz$ is the essential remaining case for an analytic classification of the restricted model, 
and provides candidate functions for future impossibility arguments.

\section{Unique-Preimage Impossibility}

In this section, we restrict the general model of Section~\ref{subsec:psqm-model} as follows. 
Each player holds exactly one qubit of the fixed shared \(k\)-qubit state \(\sigma\), 
the message register \(\mathcal M_i\) is that same qubit, and, conditioned on the shared random value \(r\in\mathcal R\) and the input bit \(x_i\), 
the local encoding channel is conjugation by a unitary \[ U_i(x_i,r)\in U(2). \] 
Thus 
\[
 \rho_{x,r} = \left( \bigotimes_{i=1}^k U_i(x_i,r) \right) \sigma \left( \bigotimes_{i=1}^k U_i(x_i,r) \right)^\dagger. 
\] 
We refer to this restriction as the \emph{one-qubit local-unitary model}.

Let $k\geq3$ and let $f:\{0,1\}^k\to\{0,1\}$. 
The players share an arbitrary mixed state $\sigma\in\mathcal D((\mathbb C^2)^{\otimes k})$ and arbitrary input-independent classical shared randomness with finite support \(\mathcal R\). 
For every player \(i\), input bit \(b\), and \(r\in\mathcal R\), the local encoder is an arbitrary unitary $U_i(b,r)\in U(2)$. 
All \(k\) qubits are sent to the referee.

We now state the main impossibility theorem.
 
\begin{theorem}[local-unitary unique-preimage impossibility]\label{thm:impossibility}
If
\[
 \min\{|f^{-1}(0)|,|f^{-1}(1)|\}=1,
\]
then no protocol in the one-qubit local-unitary model has both perfect privacy and perfect correctness.
\end{theorem}

Theorem~\ref{thm:impossibility} applies to $k$-bit AND, OR, NAND, NOR, every Boolean point function $x\mapsto\mathbf1[x=a]$, and their complements.

The proof is divided into two steps. 
Section~\ref{subsec:rank-lb} derives a rank lower bound from perfect correctness and Lemma~\ref{lem:rigidity} (zero-expectation qubit-unitary structure). 
Section~\ref{subsec:rank-ub} uses perfect privacy to obtain a contradictory rank upper bound. 

After complementing the output, assume that $f(a)=1$ for a unique $a\in\{0,1\}^k$.  
For $S\subseteq[k]$, let $a^{(S)}$ be obtained by flipping exactly the positions in $S$.  
Then $a^{(\varnothing)}=a$ and every nonempty $S$ yields output zero. 
Let $\varrho_b$ be the global state received by the referee when $b$ is output in the assumed protocol 
with perfect correctness and perfect privacy.

\subsection{Rank lower bound and purification forced by correctness}\label{subsec:rank-lb}

Take a spectral decomposition
\[
 \sigma=\sum_c\lambda_c\proj{\psi_c},
 \qquad \lambda_c>0.
\]
Fix a spectral component \(c\) and \(r\in\mathcal R\). 
Define
\[
 W_r=\bigotimes_{i=1}^k U_i(a_i,r),
 \qquad
 V_i^{(r)}=U_i(a_i,r)^{\dagger}U_i(1-a_i,r),
\]
where $a_i$ is the $i$-th bit of $a$. 
The pure component for input $a^{(S)}$ is
\[
 W_r\left(\bigotimes_{i\in S}V_i^{(r)}\right)\ket{\psi_c}.
\]
Lemma~\ref{lem:support} and perfect correctness imply
\begin{equation}\label{eq:allSzero}
 \bra{\psi_c}
 \left(\bigotimes_{i\in S}V_i^{(r)}\right)
 \ket{\psi_c}=0
 \qquad(\varnothing\neq S\subseteq[k]).
\end{equation}
Let $\tau_{i,c}:=\operatorname{Tr}_{[k]\setminus\{i\}} (\ket{\psi_c}\bra{\psi_c})$ be the reduced density operator of the $i$-th qubit of $\proj{\psi_c}$.  
Taking \(S=\{i\}\) in~\eqref{eq:allSzero} gives 
\[
 0 = \bra{\psi_c} \left( V_i^{(r)}\otimes I_{[k]\setminus\{i\}} \right) \ket{\psi_c}
 = \operatorname{Tr}\!\left( \tau_{i,c}V_i^{(r)} \right). 
\]
Applying Lemma~\ref{lem:rigidity}, we may rephase \(V_i^{(r)}\), 
which changes every subset-flip vector only by a global phase, 
so that 
\[
 (V_i^{(r)})^{\dagger}=V_i^{(r)}, \qquad
 (V_i^{(r)})^2=I.
\]
For nonempty $S$, put
\[
 \ket{\phi_S}=\left(\bigotimes_{i\in S}V_i^{(r)}\right)\ket{\psi_c}.
\]
If $S\neq T$, then
\[
 \langle\phi_S\mid\phi_T\rangle
 =\bra{\psi_c}
 \left(\bigotimes_{i\in S\triangle T}V_i^{(r)}\right)
 \ket{\psi_c}=0
\]
by \eqref{eq:allSzero}.  
Since \(W_r\) is unitary, the vectors 
\[ W_r\ket{\phi_S}, \qquad \varnothing\ne S\subseteq[k], \] are also mutually orthonormal. 
Each of them is a pure-state component of the received state for the rejecting input \(a^{(S)}\), 
and hence belongs to \(\operatorname{supp}(\varrho_0)\) by Lemma~\ref{lem:support}. 
Therefore, \begin{equation}
 \rank(\varrho_0)\geq2^k-1.
 \label{eq:ranklower}
\end{equation}
The total space has dimension $2^k$, and thus orthogonality of the output supports forces
\begin{equation}
 \rank(\varrho_1)=1.
 \label{eq:rankone}
\end{equation}
But
\[
 \varrho_1=\sum_{r,c}p_r\lambda_cW_r\proj{\psi_c}W_r^{\dagger}.
\]
Every positive component must lie in the same one-dimensional support.  
Every vector \(W_r\ket{\psi_c}\) belongs to the common one-dimensional support of \(\varrho_1\). 
On the other hand, for every fixed \(r\), the unitary \(W_r\) preserves the orthogonality of the spectral vectors: 
\[ \langle\psi_c|W_r^\dagger W_r|\psi_{c'}\rangle = \langle\psi_c|\psi_{c'}\rangle = 0 \qquad (c\ne c'). \] 
Hence \(\sigma\) can have only one positive-eigenvalue spectral component. Therefore 
\[ \operatorname{rank}(\sigma)=1, 
\] 
and the input-independent shared state is of the form 
\[ \sigma=\ket{\psi}\!\bra{\psi} \] 
for a fixed \(k\)-qubit unit vector \(\ket{\psi}\), independent of \(r\).

\subsection{Rank upper bound by privacy}\label{subsec:rank-ub}

Equation~\eqref{eq:rankone} implies that all vectors \[ W_r\ket{\psi}, \qquad r\in\mathcal R, \] lie on the same ray. 
Hence there exist a unit vector \(\ket{\eta}\) and phases \(e^{i\theta_r}\), \(r\in\mathcal R\), such that 
\[ W_r\ket{\psi} = e^{i\theta_r}\ket{\eta} \] for every \(r\in\mathcal R\). 

Choose any \(r_0\in\mathcal R\). 
Apply the fixed product unitary \(W_{r_0}^{\dagger}\) to every received state as a common change of coordinates, 
and redefine \[ W'_r := W_{r_0}^{\dagger}W_r. \] 
Since \(W_{r_0}\ket{\psi}\) and \(\ket{\eta}\) lie on the same ray, 
this common coordinate change gives 
\[ W'_r\ket{\psi} = e^{i\theta'_r}\ket{\psi} \] 
for every \(r\in\mathcal R\), after absorbing an \(r\)-independent phase into the phases \(e^{i\theta'_r}\). 
After dropping the primes, we may therefore assume that 
\begin{equation} 
W_r\ket{\psi} = e^{i\theta_r}\ket{\psi} \label{eq:rayfix} 
\end{equation} 
for every $r\in\mathcal R$.

Let
\[
 \tau_i=\Tr_{[k]\setminus\{i\}}\proj\psi
\]
be the reduced density operator of the $i$-th qubit of $\proj\psi$. 
Define
\[
 \mathcal L_i=\{M\in \mathcal L(\C^2):\Tr(\tau_iM)=0\}. 
\]
The functional $f_i: M\mapsto\Tr(\tau_iM)$ is nonzero because $\Tr(\tau_iI)=1$. 
This means the dimension of the image of $f_i$ is $1$.  
Since $\mathcal L(\C^2)$ has complex dimension four and $\mathcal L_i=\ker f_i$, 
\[
 \dim \mathcal L_i=3.
\]
Define
\[
 \mathcal A_i=\{(M_i\otimes I_{[k]\setminus\{i\}})\ket\psi:M\in\mathcal L_i\},
\]
where $M_i\otimes I_{[k]\setminus\{i\}}$ denotes the operator that acts as $M$ on the $i$-th qubit and as identity on every other qubit. 
Since $\mathcal A_i$ is the image of the linear map $M\mapsto (M_i\otimes I_{[k]\setminus\{i\}})\ket\psi$,  
\[
 \dim \mathcal A_i\leq \dim \mathcal L_i \leq3.
\]
Taking the partial trace of~\eqref{eq:rayfix} over all qubits except the \(i\)th one gives 
\[
 U_i(a_i,r)\tau_i U_i(a_i,r)^\dagger=\tau_i. 
\] 
Therefore conjugation by \(U_i(a_i,r)\) preserves \(\mathcal L_i\). 
Consequently, \(W_r\mathcal A_i=\mathcal A_i\). 

For the input $a^{(\{i\})}$ obtained by flipping only coordinate \(i\), 
Equation~\eqref{eq:allSzero} with \(S=\{i\}\) gives 
\[
 0 
= \bra{\psi} (V_i^{(r)}\otimes I) \ket{\psi} = \operatorname{Tr}(\tau_iV_i^{(r)}).
\] 
Hence \(V_i^{(r)}\in \mathcal L_i\).
Since 
\[
 \bigl(V_i^{(r)}\otimes I\bigr)\ket{\psi} \in\mathcal A_i \] and $W_r\mathcal A_i=\mathcal A_i$, 
we have 
\[ 
W_r \bigl(V_i^{(r)}\otimes I\bigr) \ket{\psi} \in\mathcal A_i 
\] 
for every \(r\in\mathcal R\). 
Thus every pure-state component of \(\rho_{a^{(\{i\})}}\) (the global state received by the referee when the input is $a^{(\{i\})}$) lies in \(\mathcal A_i\).
Consequently, 
\[ 
\operatorname{supp} \left(\rho_{ a^{(\{i\})}}\right) \subseteq \mathcal A_i. 
\] 
Since \(a^{(\{i\})}\) has $f( a^{(\{i\})} )=0$, perfect privacy gives
\[
 \supp(\varrho_0)=\supp( \rho_{a^{ ( \{ i \} ) }} ) \subseteq \mathcal A_i,
\]
and thus 
\begin{equation}
 \rank(\varrho_0)\leq3.
 \label{eq:rankupper}
\end{equation}
Combining \eqref{eq:ranklower} and \eqref{eq:rankupper} gives
\[
 2^k-1\leq\rank(\varrho_0)\leq3,
\]
a contradiction for every $k\geq3$. This proves Theorem~\ref{thm:impossibility}.


Theorem~\ref{thm:impossibility} allows any shared $k$-qubit mixed state and any amount of classical shared randomness.  
It also permits each player to possess an arbitrary untouched auxiliary system: if the input-dependent operation is restricted to $U_i(x_i,r)$ on a designated message qubit and the auxiliary system is neither acted upon nor sent, tracing out all auxiliaries reduces the setting to an arbitrary mixed state on the $k$ message qubits. 
However, the theorem is specific to qubits and local unitary encodings. 
It does not directly cover local operations that couple the message qubit to an auxiliary system, arbitrary CPTP encodings, 
protocols in which each player transmits more than one qubit, or functions for which both \(f^{-1}(0)\) and \(f^{-1}(1)\) contain at least two inputs.

\section{One-Qudit Protocols for Multiparty AND} \label{sec:qudit-and} 

The impossibility theorem of the preceding section is specific to one transmitted qubit per player. 
In this section, we show that the obstruction can be overcome by increasing the local message dimension. 
We first give a general quotient-label construction for \(\operatorname{AND}_k\). 
For every prime \(p\) satisfying $k\le 2(p-1)$, the construction uses one \(p\)-dimensional qudit per player. 
We then specialize the protocol to \(k=p=3\), obtaining a one-qutrit protocol for three-bit AND with communication \(3\log_2 3\). 
Combined with the \(6\)-bit classical lower bound of Shinagawa and Nuida~\cite{ShinagawaNuida2023}, 
this gives a strict quantum communication advantage. 
The direct three-party specialization uses \(144\) values of shared randomness. 
A three-party-specific optimization reduces the shared-randomness support to \(48\) values without changing the communication cost. 
The optimized construction is given in Appendix~\ref{app:qutrit-randomness-compression}. 

Throughout this section, all vector and matrix operations are over \(\mathbb F_p\), 
where all vectors in \(\mathbb F_p^2\) are regarded as column vectors.

\subsection{Generalized GHZ labels} \label{subsec:generalized-ghz-labels} 

We use the standard \(p\)-dimensional Weyl operators (see, e.g.,~\cite{Wat18}).

Let \(k\ge3\), and let \(p\) be a prime satisfying $k\le 2(p-1)$. 
Let $\omega=e^{2\pi i/p}$, and define the \(p\)-dimensional Weyl operators by
\[ 
 X^a\ket{j} = \ket{j+a}, \qquad Z^b\ket{j} = \omega^{bj}\ket{j}, \qquad a,b,j\in\mathbb F_p. 
\] 
The players share the \(k\)-partite \(p\)-dimensional GHZ state 
\[ 
 \ket{G_{p,k}} = \frac{1}{\sqrt p} \sum_{j\in\mathbb F_p} \ket{j}^{\otimes k}. 
\] 
For $u_2,\ldots,u_k,w\in\mathbb F_p$, define 
\[ 
 \ket{\Phi_{u_2,\ldots,u_k,w}} = \left( I \otimes X^{u_2} \otimes\cdots\otimes X^{u_{k-1}} \otimes X^{u_k}Z^w \right) \ket{G_{p,k}}. 
\] 
These \(p^k\) states form an orthonormal basis, which we call the generalized GHZ basis. 
We call $(u_2,\ldots,u_k,w)$ the GHZ label of $\ket{\Phi_{u_2,\ldots,u_k,w}}$. 

\begin{lemma}[Multiparty Weyl-to-GHZ-label rule] \label{lem:weighted-and-ghz-label} 
Suppose that player \(i\) applies \[ X^{a_i}Z^{b_i}, \qquad a_i,b_i\in\mathbb F_p. \] 
The resulting state is, up to global phase, 
$\ket{\Phi_{u_2,\ldots,u_k,w}}$, 
where 
\[ 
 u_i=a_i-a_1 \quad(2\le i\le k), \qquad w=\sum_{i=1}^k b_i. 
\] 
\end{lemma} 

\begin{proof} For every \(c\in\mathbb F_p\), 
\[
 (X^c)^{\otimes k}\ket{G_{p,k}} = \ket{G_{p,k}}. 
\] 
Hence the common shift \(a_1\) can be removed, leaving the relative shifts \(a_i-a_1\), \(2\le i\le k\). 
The phase operators contribute the total phase exponent \(\sum_i b_i\). 
Commuting \(X\)- and \(Z\)-operators may introduce a power of \(\omega\), but this is only a global phase. 
\end{proof}

\subsection{Our protocol} \label{subsec:weighted-quotient-construction} 

In this subsection, we describe our PSQM protocol. 

We first choose nonzero weights $t_2,\ldots,t_k\in\mathbb F_p^\times$ such that 
\begin{equation}  \label{eq:weighted-and-total-weight} 
T:=\sum_{i=2}^k t_i\ne0. 
\end{equation}

\begin{lemma}[Existence of suitable weights] \label{lem:weighted-and-weights} 
There exist $t_2,\ldots,t_k\in\mathbb F_p^\times$ satisfying~\eqref{eq:weighted-and-total-weight}. 
\end{lemma} 

\begin{proof} 
The condition $k\leq 2(p-1)$ and \(k\ge3\) implies \(p\ge3\). 
First set $t_2=\cdots=t_k=1$. 
If \(k-1\ne0\) in \(\mathbb F_p\), these weights give $T=k-1\ne0$. 
Suppose instead that \(k-1=0\) in \(\mathbb F_p\). Set $t_2=2$ and $t_3=\cdots=t_k=1$. 
All these weights are nonzero because \(p\ge3\), and $T=2+(k-2)=k=1$ in \(\mathbb F_p\). 
Thus \(T\ne0\). 
\end{proof} 

Fix once and for all publicly known weights $t_2,\ldots,t_k\in\mathbb F_p^\times$ satisfying~\eqref{eq:weighted-and-total-weight}. 
These weights are deterministic parameters of the protocol and are not part of the shared randomness. 
Define $\pi_{\mathbf t}: \mathbb F_p^k \longrightarrow \mathbb F_p^2$ by 
\[ 
\pi_{\mathbf t}(u_2,\ldots,u_k,w) = \left( \sum_{i=2}^k t_i u_i, w \right). 
\] 
Since \(t_2\ne0\), the map \(\pi_{\mathbf t}\) is surjective. 
Its kernel 
\[ 
K_{\mathbf t} := \ker\pi_{\mathbf t} 
\] 
has dimension \(k-2\), and hence 
\[ 
|K_{\mathbf t}|=p^{k-2}. 
\] 
For a GHZ label \[ \lambda=(u_2,\ldots,u_k,w)\in\mathbb F_p^k, \] we call \[ \pi_t(\lambda)\in\mathbb F_p^2 \] its \emph{quotient label}.

We next construct vectors governing the local input changes. 

\begin{lemma}[Zero-sum-free sequence] \label{lem:weighted-and-zero-sum-free} 
Under the condition~$k\leq 2(p-1)$, there exist vectors $s_1^{(0)},\ldots,s_k^{(0)} \in\mathbb F_p^2$ such that 
$\sum_{i\in S}s_i^{(0)}\ne0$ for every nonempty $S\subseteq[k]$. 
\end{lemma} 

\begin{proof} 
Let $e_1=(1,0)^{\mathsf T}$ and $e_2=(0,1)^{\mathsf T}$ be the standard basis of \(\mathbb F_p^2\). 
Form the sequence 
\[ 
\underbrace{e_1,\ldots,e_1}_{p-1}, \underbrace{e_2,\ldots,e_2}_{p-1}. 
\] 
Its length is \(2(p-1)\), and thus the condition $k\leq 2(p-1)$ allows us to retain the first \(k\) terms and denote them by 
\[ 
s_1^{(0)},\ldots,s_k^{(0)}. 
\] 
For a nonempty subset \(S\subseteq[k]\), 
write 
\[
 \sum_{i\in S}s_i^{(0)} = c_1e_1+c_2e_2, 
\] 
where \(0\le c_1,c_2\le p-1\). 
Since \(S\) is nonempty, \(c_1+c_2>0\), so at least one of \(c_1,c_2\) lies in \(\{1,\ldots,p-1\}\) and is nonzero modulo \(p\). 
Therefore the sum is nonzero in \(\mathbb F_p^2\).
\end{proof} 

Let $\operatorname{GL}(2,p)$ denote the group of invertible $2\times 2$ matrices over $\mathbb F_p$. 
The players share independent uniform random variables $M\in\operatorname{GL}(2,p)$ and $\tau\in K_{\mathbf t}$. 
They define 
\begin{equation}\label{eq:sec6-si} 
s_i = Ms_i^{(0)} = (\xi_i,\beta_i)^{\mathsf T} \qquad(i\in[k]). 
\end{equation} 
Writing $\tau=(\tau_2,\ldots,\tau_k,0)$, the kernel condition is 
\[ 
\sum_{i=2}^k t_i\tau_i=0. 
\] 
Define 
\begin{equation}\label{eq:sec6-h}
h := -\sum_{i=1}^k s_i = (h_1,h_2)^{\mathsf T}. 
\end{equation} 
Choose the GHZ-label representative 
\[ 
g(h) = \left( t_2^{-1}h_1, 0,\ldots,0, h_2 \right) \in\mathbb F_p^k. 
\] 
Then 
\[ 
\pi_{\mathbf t}(g(h))=h. 
\] 
To generate the input-independent GHZ label \(g(h)+\tau\), 
player \(1\) applies \(Z^{h_2}\), player \(2\) applies $X^{t_2^{-1}h_1+\tau_2}$, and player \(i\ge3\) applies \(X^{\tau_i}\). 

For the input-dependent part, player \(1\), on input \(x_1\in\{0,1\}\), applies 
\begin{equation} \label{eq:weighted-and-player-one-encoding}
X^{-x_1T^{-1}\xi_1}Z^{x_1\beta_1}, 
\end{equation} 
and player \(i\ge2\), on input \(x_i\in\{0,1\}\), applies 
\begin{equation} \label{eq:weighted-and-other-player-encoding} 
X^{x_it_i^{-1}\xi_i}Z^{x_i\beta_i}. 
\end{equation}
These operations are composed with the input-independent operations above. 
The order of the Weyl factors affects only a global phase. 
Let \[ L_{M,\tau}(x) \in\mathbb F_p^k, \qquad x=(x_1,\ldots,x_k)\in\{0,1\}^k, \] denote the resulting GHZ label. 

\paragraph{Protocol summary.} 
The public parameters are the prime \(p\), the zero-sum-free sequence $s_1^{(0)},\ldots,s_k^{(0)}$,  
the weights $t_2,\ldots,t_k$, the quotient map \(\pi_{\mathbf t}\), and its kernel \(K_{\mathbf t}\). 
For each execution, the players share independent uniform $M\in\operatorname{GL}(2,p)$ and $\tau\in K_{\mathbf t}$. 
They first compute $s_i$ and $h$ in (\ref{eq:sec6-si}) and (\ref{eq:sec6-h}). 
They then apply the local operations that generate the input-independent GHZ label \(g(h)+\tau\), 
compose them with the input-dependent operations in \eqref{eq:weighted-and-player-one-encoding} and~\eqref{eq:weighted-and-other-player-encoding}, 
and transmit one \(p\)-dimensional qudit each. 
The referee measures in the generalized GHZ basis and outputs \(1\) if and only if the observed GHZ label belongs to \(K_{\mathbf t}\).

\subsection{Correctness and privacy} \label{subsec:weighted-and-correctness-privacy} 

\begin{lemma}[Quotient label formula] \label{lem:weighted-and-quotient-formula} For every \(x\in\{0,1\}^k\), 
\begin{equation} \label{eq:weighted-and-quotient-formula}
\pi_{\mathbf t}(L_{M,\tau}(x)) = -\sum_{i:x_i=0}s_i. 
\end{equation}
\end{lemma} 

\begin{proof} 
The input-independent GHZ label has the quotient label 
\[ 
\pi_{\mathbf t}(g(h)+\tau) = \pi_{\mathbf t}(g(h)) + \pi_{\mathbf t}(\tau) = h = -\sum_{i=1}^k s_i. 
\]

For player \(i\ge2\), increasing \(x_i\) from \(0\) to \(1\) changes the \(u_i\)-component of the GHZ label by \(t_i^{-1}\xi_i\) and its last component \(w\) by \(\beta_i\). 
Hence the quotient label changes by \[ \bigl(t_i(t_i^{-1}\xi_i),\beta_i\bigr) = (\xi_i,\beta_i) = s_i. \]

For player \(1\), increasing \(x_1\) from \(0\) to \(1\) changes the first player's \(X\)-exponent \(a_1\) by $-T^{-1}\xi_1$.  
By Lemma~\ref{lem:weighted-and-ghz-label}, 
the \(u_i\)-component of the GHZ label is 
\[
 u_i=a_i-a_1, \qquad 2\le i\le k. 
\] 
Since the exponents \(a_i\) for \(i\ge2\) remain unchanged, every \(u_i\)-component changes by $T^{-1}\xi_1$.  
Therefore the first component of the quotient label changes by 
\[
 \sum_{i=2}^k t_iT^{-1}\xi_1 = TT^{-1}\xi_1 = \xi_1. 
\]
The last component \(w\) of the GHZ label changes by \(\beta_1\). 
Therefore, when \(x_1\) changes from \(0\) to \(1\), the quotient label changes by
 \[ (\xi_1,\beta_1) = s_1. \]

Thus changing \(x_i\) from \(0\) to \(1\) changes the quotient label by \(s_i\) for every \(i\in[k]\). 
Since the input-independent GHZ label has quotient label 
\[ -\sum_{i=1}^k s_i, \] 
we obtain 
\[
\begin{aligned} 
\pi_{\mathbf t}\bigl(L_{M,\tau}(x)\bigr) 
&= -\sum_{i=1}^k s_i + \sum_{i=1}^k x_i s_i\\ 
&= -\sum_{i:x_i=0}s_i. 
\end{aligned} 
\]
\end{proof} 

\begin{proposition}[Perfect correctness] \label{prop:weighted-and-correctness} 
For every $M\in\operatorname{GL}(2,p)$ and $\tau\in K_{\mathbf t}$, 
\[
 L_{M,\tau}(x)\in K_{\mathbf t} \quad\Longleftrightarrow\quad x=1^k. 
\] 
\end{proposition} 

\begin{proof} 
If \(x=1^k\), \eqref{eq:weighted-and-quotient-formula} gives 
\[ 
 \pi_{\mathbf t}(L_{M,\tau}(1^k))=0. 
\] 
Hence \(L_{M,\tau}(1^k)\in K_{\mathbf t}\). 
Now suppose that \(x\ne1^k\), and define the zero-position set of $x$: 
\[ 
S_x:=\{i\in [k]: x_i=0\}. 
\] 
Then \(S_x\) is nonempty, and 
\[ 
\sum_{i\in S_x}s_i = M\left( \sum_{i\in S_x}s_i^{(0)}\right). 
\] 
The inner sum is nonzero by Lemma~\ref{lem:weighted-and-zero-sum-free}. 
Since \(M\) is invertible, 
\[ 
\sum_{i\in S_x}s_i\ne0. 
\] 
Equation~\eqref{eq:weighted-and-quotient-formula} therefore gives 
\[ 
\pi_{\mathbf t}(L_{M,\tau}(x))\ne0, 
\] 
and hence \(L_{M,\tau}(x)\notin K_{\mathbf t}\). 
\end{proof} 

\begin{lemma}[Uniform nonzero quotient labels] \label{lem:weighted-and-uniform-quotient} 
Fix a nonempty subset \(S\subseteq[k]\). If \(M\) is uniform on \(\operatorname{GL}(2,p)\), 
then $\sum_{i\in S}s_i$ is uniform on $\mathbb F_p^2\setminus\{0\}$.  
\end{lemma} 

\begin{proof} 
Set 
\[
 v_S:=\sum_{i\in S}s_i^{(0)}. 
\] 
Lemma~\ref{lem:weighted-and-zero-sum-free} gives \(v_S\ne0\), and 
\[
 \sum_{i\in S}s_i=Mv_S. 
\] 
The group \(\operatorname{GL}(2,p)\) acts transitively on \(\mathbb F_p^2\setminus\{0\}\). 
More explicitly, for any nonzero \(y,z\), choose \(A\in GL(2,p)\) with \(Ay=z\). 
Then \(M\mapsto AM\) is a bijection between the matrices satisfying \(Mv_S=y\) and those satisfying \(Mv_S=z\). 
Thus every nonzero vector has the same number of preimages, and \(Mv_S\) is uniform on the nonzero vectors.
\end{proof} 

\begin{proposition}[Perfect privacy] \label{prop:weighted-and-privacy} 
After averaging over the shared randomness \(M\) and \(\tau\), 
the state received by the referee depends only on the value of $\operatorname{AND}_k$. 
\end{proposition} 

\begin{proof} 
For each \(y\in\mathbb F_p^2\), choose a GHZ label \(\lambda_y\in\mathbb F_p^k\) satisfying 
\[
 \pi_{\mathbf t}(\lambda_y)=y. 
\] 
The set of GHZ labels with the quotient label \(y\) is the coset 
\[
 C_y := \lambda_y+K_{\mathbf t} = \pi_{\mathbf t}^{-1}(y). 
\] 
Every such coset has cardinality 
\[
 |C_y|=|K_{\mathbf t}|=p^{k-2}. 
\] 
Fix \(M\) and an input \(x\). 
The GHZ label depends additively on \(\tau\). 
Since \(\tau\) is uniform on \(K_{\mathbf t}\), 
conditioned on the quotient label
\[
 y=\pi_{\mathbf t}(L_{M,\tau}(x)), 
\] 
the GHZ label is uniform on \(C_y\). 

For the accepting input \(x=1^k\), the quotient label is \(0\), 
and thus, the averaged received state is 
\[
 \varrho_1 = \frac{1}{p^{k-2}} \sum_{\lambda\in K_{\mathbf t}} \ket{\Phi_\lambda}\!\bra{\Phi_\lambda}. 
\] 

Now fix a rejecting input \(x\ne1^k\). Its zero-position set \(S_x\) is nonempty, 
and 
\[
 \pi_{\mathbf t}(L_{M,\tau}(x)) = -\sum_{i\in S_x}s_i. 
\] 
By Lemma~\ref{lem:weighted-and-uniform-quotient}, this quotient label is uniform on $\mathbb F_p^2\setminus\{0\}$, independently of \(S_x\). 
Conditioned on each nonzero quotient label, the kernel shift makes the GHZ label uniform on the corresponding coset. 
Hence every rejecting input induces 
\[ 
\varrho_0 = \frac{1}{(p^2-1)p^{k-2}} \sum_{\substack{y\in\mathbb F_p^2\\y\ne0}} \sum_{\lambda\in C_y} \ket{\Phi_\lambda}\!\bra{\Phi_\lambda} 
= \frac{1}{p^k-p^{k-2}} \sum_{\lambda\notin K_{\mathbf t}} \ket{\Phi_\lambda}\!\bra{\Phi_\lambda}. 
\] 
This state is independent of the rejecting input. 
\end{proof}

\subsection{Consequences} \label{subsec:weighted-and-consequences} 

We can now state the resulting PSQM upper bound for \(\operatorname{AND}_k\).

\begin{theorem}[One-qudit protocols for multiparty AND] \label{thm:weighted-qudit-and} 
Let \(k\ge3\), and let \(p\) be a prime satisfying $k\le2(p-1)$.  
Then there is a perfectly correct and perfectly private entanglement-assisted PSQM protocol for \(\operatorname{AND}_k\) in which the players share one \(k\)-partite \(p\)-dimensional GHZ state and every player transmits one \(p\)-dimensional qudit. 
Consequently, 
\[
 Q^{\mathrm{PSQM},*}(\operatorname{AND}_k) \le k\log_2 p. 
\] 
\end{theorem} 

\begin{proof} 
The protocol is defined in Section~\ref{subsec:weighted-quotient-construction}. 
Perfect correctness follows from Proposition~\ref{prop:weighted-and-correctness}, 
and perfect privacy follows from Proposition~\ref{prop:weighted-and-privacy}. 
Since each player transmits one \(p\)-dimensional register, 
the communication cost is $k\log_2 p$.  
\end{proof} 

\begin{corollary} \label{cor:weighted-and-all-k} 
For every \(k\ge3\), there exists a prime \(p\le k\) such that $k\le2(p-1)$.  
Consequently, 
\[
 Q^{\mathrm{PSQM},*}(\operatorname{AND}_k) \le k\log_2 p \le k\log_2 k.
\] 
\end{corollary} 

\begin{proof} Let $n=\left\lceil\frac{k}{2}\right\rceil$. 
Since \(k\ge3\), we have \(n>1\). By Bertrand's postulate, there is a prime \(p\) satisfying $n<p<2n$. 
Since \(p\) and \(n\) are integers, $p-1\ge n = \left\lceil\frac{k}{2}\right\rceil$, and hence $k\le2(p-1)$. 
If \(k\) is even, then \(2n=k\), so \(p<k\). If \(k\) is odd, then \(2n=k+1\), so \(p<k+1\), and hence \(p\le k\). 
The result follows from Theorem~\ref{thm:weighted-qudit-and}. 
\end{proof} 

The classical construction for \(\operatorname{AND}_k\) in \cite{FKN94} chooses a prime \(\ell>k\) 
and uses message alphabet size \(\ell\) for every player, giving communication $k\log_2\ell$, 
which remains the best general-\(k\) classical upper bound for AND known to us. 
Corollary~\ref{cor:weighted-and-all-k} gives a one-shot quantum protocol with local dimension \(p\le k\), and therefore $k\log_2p < k\log_2\ell$. 
Thus our quantum construction improves on the achievable communication cost in \cite{FKN94} for every \(k\ge3\). 
This compares achievable upper bounds and does not by itself establish a separation from the optimal classical PSM complexity 
unless a sufficiently strong classical lower bound is known. 

\paragraph{The three-party qutrit protocol}

Taking \(k=3\) in Corollary~\ref{cor:weighted-and-all-k} gives the following special case.   

\begin{corollary}[One-qutrit protocol for three-bit AND] \label{cor:qutrit-and} 
There is a perfectly correct and perfectly private entanglement-assisted PSQM protocol for \(\operatorname{AND}_3\) using one shared three-qutrit GHZ state and one transmitted qutrit per player. 
Consequently, 
\[ 
Q^{\mathrm{PSQM},*}(\operatorname{AND}_3) \le 3\log_2 3.
\] 
\end{corollary} 

The direct specialization uses independent uniform shared randomness $M\in\operatorname{GL}(2,3)$ and $\tau\in K_{\mathbf t}$.  
Since $|\operatorname{GL}(2,3)| = (3^2-1)(3^2-3) = 48$ and $|K_{\mathbf t}|=3$, this protocol has \(144\) values of shared randomness.  
For three players, however, the independent kernel randomizer can be removed. 
Appendix~\ref{app:qutrit-randomness-compression} gives another perfectly correct and perfectly private protocol with the same communication cost 
whose shared randomness is a single uniform $M\in\operatorname{GL}(2,3)$.  
Thus the optimized implementation uses only \(48\) shared-randomness values. 

Shinagawa and Nuida~\cite{ShinagawaNuida2023} showed that the message-alphabet sizes of any classical PSM protocol for \(\operatorname{AND}_3\), 
up to permutation and componentwise enlargement, must dominate one of $(3,6,6),(4,4,4),(4,4,5),(4,4,6),(4,4,7),(4,5,5),(4,5,6)$ and $(5,5,5)$.  
The smallest product among these tuples is $4\cdot4\cdot4=64$.  
Hence every classical PSM protocol for \(\operatorname{AND}_3\) communicates at least $\log_2 64=6$ bits. 
In contrast, Corollary~\ref{cor:qutrit-and} gives an entanglement-assisted PSQM protocol communicating $3\log_2 3 = \log_2 27 < 6$ qubits.

Corollary~\ref{cor:qutrit-and} complements Theorem~\ref{thm:impossibility}. 
Theorem~\ref{thm:impossibility} rules out a protocol for \(\operatorname{AND}_3\) in the one-qubit local-unitary model, 
even with an arbitrary shared three-qubit state and arbitrary shared randomness. 
Corollary~\ref{cor:qutrit-and} shows that one qutrit per player is sufficient. 
Thus, within the symmetric one-system-per-player local-unitary setting, local dimension three is the smallest possible local dimension for \(\operatorname{AND}_3\). 

Lu, Yao, and Jafar~\cite{LYJ23} constructed a non-one-shot protocol for $\operatorname{AND}_k$ 
by applying the protocol in \cite{FKN94} separately to each AND instance and jointly computing 
the resulting masked sums using a capacity-achieving secure $k$-sum protocol. 
Each AND instance is reduced to a secure sum over $\mathbb F_p$, where $k<p<2k$. 
The resulting {amortized cost} is $\frac k2\log_2 p$ qubits per computation. 
For \(k=3\), this is $\frac32\log_2 5$. 
This value is an asymptotic per-instance cost obtained by jointly encoding multiple product instances. 
In contrast, Corollary~\ref{cor:qutrit-and} is a one-shot construction for a single instance of \(\operatorname{AND}_3\).

\section{Conclusion}

The communication complexity of natural functions remains much less understood in PSQM than quantum separations for specially designed functions. 
Our results show that familiar total functions already exhibit both genuine quantum advantages and strong structural limitations. 

Prior entanglement reduces communication for multiparty equality and three-bit majority. 
In particular, majority requires only three communicated qubits, compared with the optimal classical cost of six bits. 
Together with the protocols for three-bit equality and a wider family of Boolean functions, 
these results show that quantum advantage in PSQM is not confined to artificial promise problems. 

The contrast with AND shows that the structure of the function is equally important. Functions with a unique accepting or rejecting input are incompatible with perfect correctness and perfect privacy in the one-qubit local-unitary model, regardless of the shared \(k\)-qubit state and shared randomness. This is a general structural obstruction, not a failure of a particular GHZ construction. 

Increasing the local dimension overcomes the obstruction. Our quotient-label construction gives one-shot protocols for multiparty AND using one qudit per player with dimension at most the number of players. For three-bit AND, one qutrit per player is sufficient, whereas one qubit per player is impossible in the local-unitary model. Thus the benefit of entanglement in PSQM is governed jointly by the structure of the function and the local dimension available to the players.

Several questions remain open. 

First, the unique-preimage theorem is specific to unitary operations on a designated message qubit. 
Arbitrary local CPTP encodings, input-dependent interactions with auxiliary systems, and protocols in which each player transmits more than one qubit may evade the three-dimensional local-action bound. 
It remains open whether the one-shot quantum communication cost of \(\operatorname{AND}_3\) can be reduced below $3\log_2 3$ using asymmetric message dimensions, arbitrary local CPTP encodings, or input-dependent interactions with auxiliary systems. 
Equivalently, it is unknown whether the qutrit protocol is optimal among all entanglement-assisted PSQM protocols. 
The exact classical and quantum communication complexities of \(\operatorname{AND}_3\) also remain open, 
although the known bounds already give a strict separation. 

Second, the finite GHZ--Pauli experiment leaves \(176\) functions without feasible numerical solutions. 
The most structured unresolved cases include \(\operatorname{EXACT}_1\), \(\operatorname{EXACT}_2\), 
and the affine extensions of two-edge quadratic forms such as \(xy\oplus xz\). 
A rigorous infeasibility proof or a direct analytic converse would turn the numerical census into a complete theorem for the restricted model. 

Third, the construction for $\AEQ_{k,n}$ relies on existential multicolored sum-free families and on a reduction through generalized equality. 
It remains open to obtain uniform and efficiently computable local maps, improve the lower-order terms, or construct direct protocols that bypass the zero-sum layer. 
It would also be natural to ask which other standard multiparty predicates admit comparable entanglement-assisted reductions. 

More generally, our results suggest a program for PSQM centered on concrete functions. 
Equality and majority demonstrate that prior entanglement can reduce the communication cost of familiar total functions. 
The contrast between majority and AND shows that changing the structure of the accepting and rejecting inputs can create a strong obstruction. 
A fuller understanding of quantum advantage in secure simultaneous communication will therefore require both constructive mechanisms, 
such as combinatorial embeddings and entangled secure sums, and structural tools that explain when privacy is incompatible with local quantum encodings.

\paragraph{AI disclosure.}
Microsoft Copilot and Chat GPT were used to assist with iterative discussion, manuscript editing, and the generation and execution of code for the exploratory linear-programming and finite-search experiments. The author independently verified the final analytic constructions and takes full responsibility for the contents of the paper.

\paragraph{Acknowledgements.}
This work is supported by JSPS KAKENHI 24H00071, 24K22293, 26H02504, MEXT Q-LEAP JPMXS0120319794, and JST CREST JPMJCR24I4.

\appendix 

\section{Shared-Randomness Compression for the Three-Party Qutrit Protocol} \label{app:qutrit-randomness-compression} 

The direct specialization of the quotient-label construction to \(k=p=3\) uses independent uniform random variables $M\in\operatorname{GL}(2,3)$ 
and $\tau\in K_{\mathbf t}$, and therefore has $48\cdot3=144$ values. 
In this appendix, we give a three-party-specific optimization that removes the independent kernel randomizer and reduces the shared-randomness support to \(48\) values. 

The optimized construction preserves the communication cost \(3\log_2 3\), but produces different output-dependent density operators from those of the direct specialization. 
Instead of using an independent random shift in the kernel, it correlates the choice of a GHZ label with the uniformly random matrix \(M\), subject to the prescribed quotient label. 
The resulting mixtures over GHZ-basis states are nonuniform. 
Nevertheless, every rejecting input produces the same density operator, 
while the accepting input produces another density operator with orthogonal support. 
Thus perfect privacy and perfect correctness are preserved.

Throughout this appendix, 
all vector and matrix operations are over \(\mathbb F_3\).
We specialize the notation of Section~\ref{subsec:generalized-ghz-labels} by writing 
\[ 
\ket{G_3}:=\ket{G_{3,3}}= \frac{1}{\sqrt3} \sum_{j\in\mathbb F_3} \ket{j,j,j}
\]
and 
\[ 
\ket{\Phi_{u,v,w}} := \ket{\Phi_{u_2=u,u_3=v,w}} = \left( I\otimes X^u\otimes X^vZ^w \right)\ket{G_3}. 
\] 
By Lemma~\ref{lem:weighted-and-ghz-label}, local Weyl exponents $(a_1,b_1)$, $(a_2,b_2)$, and $(a_3,b_3)$ produce the GHZ label 
\begin{equation}\label{eq:qutrit-ghz-label-rule}
 (a_2-a_1,\ a_3-a_1,\ b_1+b_2+b_3). 
\end{equation} 
The quotient map specializes to \[ \pi(u,v,w)=(u+v,w), \] with kernel 
\[
 \ker\pi= \{(0,0,0),(1,2,0),(2,1,0)\}.
\]  
The referee decodes a GHZ label in \(\ker\pi\) as output \(1\), and every other GHZ label as output \(0\).

\subsection{The \(48\)-value construction}
Let 
\begin{equation} \label{eq:qutrit-base-triple}
s_1^{(0)}=s_2^{(0)}=(0,1)^{\mathsf T}, \qquad s_3^{(0)}=(1,2)^{\mathsf T}
\end{equation}
in \(\mathbb{F}_3^2\). 
Every nonempty subset sum of these three vectors is nonzero. 
Explicitly, 
\begin{equation} \label{eq:qutrit-base-subset-sums}
\begin{aligned} 
s_1^{(0)}&=(0,1)^{\mathsf T},& s_2^{(0)}&=(0,1)^{\mathsf T},& s_3^{(0)}&=(1,2)^{\mathsf T},\\
 s_1^{(0)}+s_2^{(0)}&=(0,2)^{\mathsf T},& s_1^{(0)}+s_3^{(0)}&=(1,0)^{\mathsf T},& s_2^{(0)}+s_3^{(0)}&=(1,0)^{\mathsf T},\\
 s_1^{(0)}+s_2^{(0)}+s_3^{(0)}&=(1,1)^{\mathsf T}. 
\end{aligned} 
\end{equation}
The players share a uniformly random matrix 
\[ 
M = \begin{pmatrix} a&b\\ c&d \end{pmatrix} \in\operatorname{GL}(2,3). 
\] 
Since $|\operatorname{GL}(2,3)|= 48$, the shared-randomness support has size \(48\).
Set 
\[
 s_i=Ms_i^{(0)} \qquad (i=1,2,3). 
\] 
Thus 
\[
 s_1=s_2=(b,d), \qquad s_3=(a+2b,c+2d). 
\] 
Players \(1,2,3\) hold the local input bits \(x,y,z\), respectively, and apply the following Weyl operators: 
\begin{equation}\label{eq:qutrit-and-encoders}
\begin{aligned}
 A_x &= X^{bx}Z^{dx},\\ B_y &= X^{a+by}Z^{dy},\\ C_z &= X^{a+2b+(a+2b)z} Z^{2c+2d+(c+2d)z}. 
\end{aligned} 
\end{equation}
The order of the \(X\)- and \(Z\)-factors affects only a global phase. 

The input-dependent parts of these operators are exactly the specialization of (\ref{eq:weighted-and-player-one-encoding}) and (\ref{eq:weighted-and-other-player-encoding}) to $k=p=3$ and $t_2=t_3=1$. 
The input-independent factors select an $M$-dependent GHZ-label representative of the required quotient label. 
Equivalently, they replace the independent uniform kernel shift of the general construction by a deterministic kernel shift correlated with $M$. 

Let 
\[
 L_M(x,y,z) = \bigl( u_M(x,y,z), v_M(x,y,z), w_M(x,y,z) \bigr) \in\mathbb{F}_3^3 
\] denote the resulting GHZ label. 
By (\ref{eq:qutrit-ghz-label-rule}), 
\begin{equation}\label{eq:qutrit-and-label-explicit}
\begin{aligned}
 u_M(x,y,z) &= a+b(y-x),\\ v_M(x,y,z) &= a+2b+(a+2b)z-bx,\\ w_M(x,y,z) &= 2c+2d+d(x+y)+(c+2d)z. 
\end{aligned}  
\end{equation}

\subsection{Correctness} 

The quotient label admits a particularly simple description. 

\begin{lemma}[Quotient label formula] \label{lem:qutrit-and-quotient} 
For every \(M\in\operatorname{GL}(2,3)\) and \((x,y,z)\in\{0,1\}^3\), 
\begin{equation} \label{eq:qutrit-and-quotient-formula}
\pi(L_M(x,y,z)) = -(1-x)s_1-(1-y)s_2-(1-z)s_3. 
\end{equation}
\end{lemma} 

\begin{proof}
The input-dependent factors are the \(k=p=3\), \(t_2=t_3=1\) specialization of the general construction, 
and thus Lemma~\ref{lem:weighted-and-quotient-formula} applies to them. 
The input-independent GHZ label \[ g_M=(a,a+2b,2c+2d) \] satisfies \[ \pi(g_M) = -(a+b,c+d) = -(s_1+s_2+s_3). \] 
The claimed formula follows.
\end{proof}

\begin{proposition}[Perfect correctness] \label{prop:qutrit-and-correctness} 
For every \(M\in\operatorname{GL}(2,3)\), 
\[ L_M(x,y,z)\in \ker\pi \quad\Longleftrightarrow\quad x=y=z=1. \] 
\end{proposition} 

\begin{proof} 
For input \(111\), \eqref{eq:qutrit-and-quotient-formula} gives 
\[ \pi(L_M(1,1,1))=0, \] 
and hence \(L_M(1,1,1)\in \ker\pi\). 
Now suppose that \((x,y,z)\ne(1,1,1)\), and define the nonempty set \[ T=\{i\in\{1,2,3\}:x_i=0\}, \] 
where \(x_1=x\), \(x_2=y\), and \(x_3=z\). 
Then \[ \pi(L_M(x,y,z)) = -\sum_{i\in T}s_i = -M\sum_{i\in T}s_i^{(0)}. \] 
The corresponding subset sum of the vectors in \eqref{eq:qutrit-base-triple} is nonzero by \eqref{eq:qutrit-base-subset-sums}. 
Since \(M\) is invertible, its image is also nonzero. Therefore \[ \pi(L_M(x,y,z))\ne0, \] so \(L_M(x,y,z)\notin \ker\pi\). 
\end{proof} 
Thus a GHZ-basis measurement followed by membership testing in \(\ker\pi\) computes \(\operatorname{AND}_3\) with perfect correctness.

\subsection{Perfect privacy} 

By Lemma~\ref{lem:qutrit-and-quotient} and the \(p=3\) case of Lemma~\ref{lem:weighted-and-uniform-quotient}, 
every rejecting input produces the uniform distribution on $\mathbb F_3^2\setminus\{0\}$ at the level of the quotient label. 
Unlike the general construction, however, the present protocol does not use an independent uniform shift in $\ker\pi$. 
Uniformity of the quotient label therefore does not by itself imply privacy of the received state. 
We must additionally verify that all rejecting inputs induce the same GHZ-label distribution. 

Define 
\[
 R_3 := \{ (1,2,1),(1,2,2),(2,1,1),(2,1,2) \},  
\] 
and 
\[
 N := \{(0,0,1),(0,0,2)\}. 
\] 
Set 
\[
 R_2 := \mathbb{F}_3^3 \setminus \bigl(\ker\pi \cup R_3\cup N\bigr). 
\] 
Then $|R_2|=18$ and $|R_3|=4$.  

\begin{lemma}[GHZ-label distributions] \label{lem:qutrit-and-label-distributions} 
Let \(M\) be uniform on \(\operatorname{GL}(2,3)\). 
For every rejecting input \((x,y,z)\ne(1,1,1)\), 
the GHZ label \(L_M(x,y,z)\) has the same distribution, given by 
\begin{equation} \label{eq:qutrit-reject-label-distribution} 
\Pr[L_M(x,y,z)=\lambda] = \begin{cases} 2/48, & \lambda\in R_2,\\ 3/48, & \lambda\in R_3,\\ 0, & \lambda\in \ker\pi\cup N. \end{cases} 
\end{equation} 
For the accepting input, 
\begin{equation} \label{eq:qutrit-accept-label-distribution}
\Pr[L_M(1,1,1)=\lambda] = \begin{cases} 12/48, & \lambda=(0,0,0),\\ 18/48, & \lambda=(1,2,0),\\ 18/48, & \lambda=(2,1,0),\\ 0, & \lambda\notin\ker\pi. \end{cases} 
\end{equation} 
\end{lemma} 

\begin{proof}
Write 
\[
 M= \begin{pmatrix} a&b\\ c&d \end{pmatrix}. 
\] 
Substituting the eight possible inputs into \eqref{eq:qutrit-and-label-explicit} reduces the assertion to counting the quadruples $(a,b,c,d)\in\mathbb{F}_3^4$ satisfying $ad-bc\ne0$.  

For input \(111\), \eqref{eq:qutrit-and-label-explicit} simplifies to \[ L_M(1,1,1)=(a,2a,0). \] 
There are \(12\) invertible matrices with \(a=0\), and \(18\) each with \(a=1\) and \(a=2\). 
This gives \eqref{eq:qutrit-accept-label-distribution}. 

For each of the other seven inputs, 
one substitutes the corresponding values into~\eqref{eq:qutrit-and-label-explicit} 
and counts the solutions of $ad-bc\ne0$, which gives (\ref{eq:qutrit-reject-label-distribution}). 

We illustrate the rejecting-input count for \(000\). 
In this case, \eqref{eq:qutrit-and-label-explicit} gives \[ L_M(0,0,0) = (a,a+2b,2c+2d). \] 
Fix a GHZ label \(\lambda=(u,v,w)\). The identity \(L_M(0,0,0)=\lambda\) is equivalent to 
\[ a=u, \qquad b=2(v-u), \qquad c=2w-d, \] where \(d\in\mathbb F_3\) is free. 
Substitution into the determinant gives \[ \det M = 2(u+v)d+(u-v)w. \] 
If \(u+v\ne0\), this is a nonconstant affine function of \(d\), 
so it vanishes for exactly one of the three values of \(d\). 
Hence the GHZ label has exactly two preimages in \(\operatorname{GL}(2,3)\). 
If \(u+v=0\), then \[ \det M=(u-v)w. \] 
This constant is nonzero exactly when \(u\ne0\) and \(w\ne0\), which gives the four GHZ labels in \(R_3\). 
Otherwise either \(w=0\), giving a GHZ label in \(\ker\pi\), or \(u=v=0\), giving a GHZ label in \(N\). 
In both cases, \(\det M\) vanishes for every value of the free parameter \(d\). 

For each of the other six rejecting inputs, 
substituting the corresponding values into \eqref{eq:qutrit-and-label-explicit} gives the same multiplicity pattern. 
More precisely, after solving the three GHZ-label equations, one free parameter remains. 
For GHZ labels in \(R_2\), 
the determinant is a nonconstant affine function of this parameter and therefore vanishes once; for GHZ labels in \(R_3\), 
it is a nonzero constant; and for GHZ labels in \(\ker\pi\cup N\), it vanishes identically. 
Hence the multiplicities are \(2\), \(3\), and \(0\), respectively, for every rejecting input.
\end{proof} 

Define the two density operators 
\[
 \varrho_0 := \frac{1}{48} \left( 2\sum_{\lambda\in R_2} \ket{\Phi_\lambda}\!\bra{\Phi_\lambda} + 3\sum_{\lambda\in R_3} \ket{\Phi_\lambda}\!\bra{\Phi_\lambda} \right) 
\] 
and 
\[
 \varrho_1 := \frac{1}{48} \left( 12\ket{\Phi_{0,0,0}}\!\bra{\Phi_{0,0,0}} + 18\ket{\Phi_{1,2,0}}\!\bra{\Phi_{1,2,0}} + 18\ket{\Phi_{2,1,0}}\!\bra{\Phi_{2,1,0}} \right). 
\] 
Lemma~\ref{lem:qutrit-and-label-distributions} gives 
\begin{equation} \label{eq:qutrit-and-output-states}
\rho_{(x,y,z)} = \begin{cases} \varrho_0, & (x,y,z)\ne(1,1,1),\\ \varrho_1, & (x,y,z)=(1,1,1), 
\end{cases} 
\end{equation} 
where $\rho_{(x,y,z)}$ is the state received by the referee when the input is $(x,y,z)$. 
Thus the joint state received by the referee depends only on the value of \(\AND\). 

We have proved the following theorem. 

\begin{theorem}[$48$-value three-party qutrit protocol] \label{thm:qutrit-and-compressed} 
There is a perfectly correct and perfectly private entanglement-assisted PSQM protocol for \(\operatorname{AND}_3\) 
in which the players share one three-qutrit GHZ state and a uniformly random matrix \(M\in\operatorname{GL}(2,3)\), 
and each player transmits one qutrit. The communication cost is \(3\log_2 3\), and the shared-randomness support has size \(48\).
\end{theorem}

\begin{proof} 
Perfect correctness follows from Proposition~\ref{prop:qutrit-and-correctness}. 
Perfect privacy follows from~\eqref{eq:qutrit-and-output-states}. 
Each player transmits one three-dimensional quantum register, so the total communication cost is $3\log_2 3$.  
\end{proof}

\end{document}